\documentclass[article]{IEEEtran}
\usepackage{fancyhdr}
\usepackage{subcaption}
\usepackage{physics}
\usepackage{amsmath,amsthm,amssymb}
\usepackage{filecontents}
\usepackage{siunitx}
\usepackage{graphicx} % Required for inserting images
\usepackage{optidef}
\usepackage{bbm}
\usepackage{amsmath}
\usepackage{amsfonts}
\usepackage{optidef}
\usepackage{algorithmic,eqparbox,array}
\usepackage{xcolor}
\usepackage{tikz}
\usepackage{hyperref}
\usetikzlibrary{decorations.pathreplacing,calc}
\usepackage[ruled,vlined]{algorithm2e}
\usepackage{cite}[sorting=none]
\newcommand{\STATEnonum}{\item[]}

\theoremstyle{definition}

\newtheorem{proposition}{Proposition}
\newtheorem{assumption}{Assumption}

\title{Conformal Lyapunov Optimization: \\Optimal Resource Allocation under Deterministic  Reliability Constraints}
\author{Francesco Binucci, Osvaldo Simeone, and Paolo Banelli \thanks{The work of Francesco Binucci and Paolo Banelli was supported by the European Union - Next Generation EU under the Italian National Recovery and Resilience Plan (NRRP), Mission 4, Component 2, Investment 1.3, CUP F83C22001690001/E83C22004640001, partnership on “Telecommunications of the Future” (PE00000001 - program “RESTART”). The work of Osvaldo Simeone was partially supported by the European Union’s Horizon Europe project CENTRIC (101096379),  by the Open Fellowships of the EPSRC (EP/W024101/1) and by the EPSRC project (EP/X011852/1).

Francesco Binucci and Paolo Banelli are with the Department of Engineering, University of Perugia, Via G. Duranti 93 06125, Perugia, Italy (email: paolo.banelli@unipg.it, francesco.binucci@dottorandi.unipg.it). Francesco Binucci is also with the Consorzio Nazionale Interuniversitario per le Telecomunicazioni (CNIT). 

Osvaldo Simeone is with the King’s Communications, Learning \& Information Processing (KCLIP) lab within the Centre for Intelligent Information Processing Systems (CIIPS), Department of Engineering, King’s College London, London WC2R 2LS, U.K. (e-mail: osvaldo.simeone@kcl.ac.uk).}}
\begin{document}

\maketitle

\begin{abstract}
This paper introduces conformal Lyapunov optimization (CLO), a novel resource allocation framework for networked systems that optimizes average long-term objectives,  while satisfying deterministic long-term reliability constraints. Unlike traditional Lyapunov optimization (LO), which addresses resource allocation tasks under average long-term constraints, CLO provides formal worst-case deterministic reliability guarantees. This is achieved by integrating the standard LO optimization framework with online conformal risk control (O-CRC), an adaptive update mechanism controlling long-term risks. The effectiveness of CLO is verified via experiments for hierarchal edge inference targeting  image segmentation tasks in a networked computing architecture. Specifically, simulation results confirm that CLO can   control reliability constraints,  measured via the false negative rate of all the segmentation decisions made in the network, while at the same time minimizing the weighted sum of energy consumption and precision loss, with the latter accounting for the rate of false positives.
\end{abstract}

\section{Introduction}
\subsection{Context and Motivation}
Dynamic resource allocation for networked systems is a well-established research area \cite{ribeiro2012optimal}, which has acquired new dimensions with the advent of mobile edge computing (MEC)  \cite{mao2017survey} in 5G networks and beyond \cite{Zhang2020dynamic}. For networks involving mobile devices with limited energy and computational resources, it is becoming increasingly important to offer computing services closer to the edge  for artificial intelligence (AI) workloads, while satisfying diverse and stringent requirements in terms of energy consumption, latency, and reliability \cite{zhou2019edge} (see Figure 1). For instance, for ultra-reliable and low-latency communications (URLLC) traffic, including autonomous driving \cite{chang2021autonomous} and Industry 4.0 \cite{rao2018impact}, timely decision-making with guaranteed reliability is paramount.

\begin{figure*}[ht]
    \centering
    \includegraphics[width=0.55\linewidth]{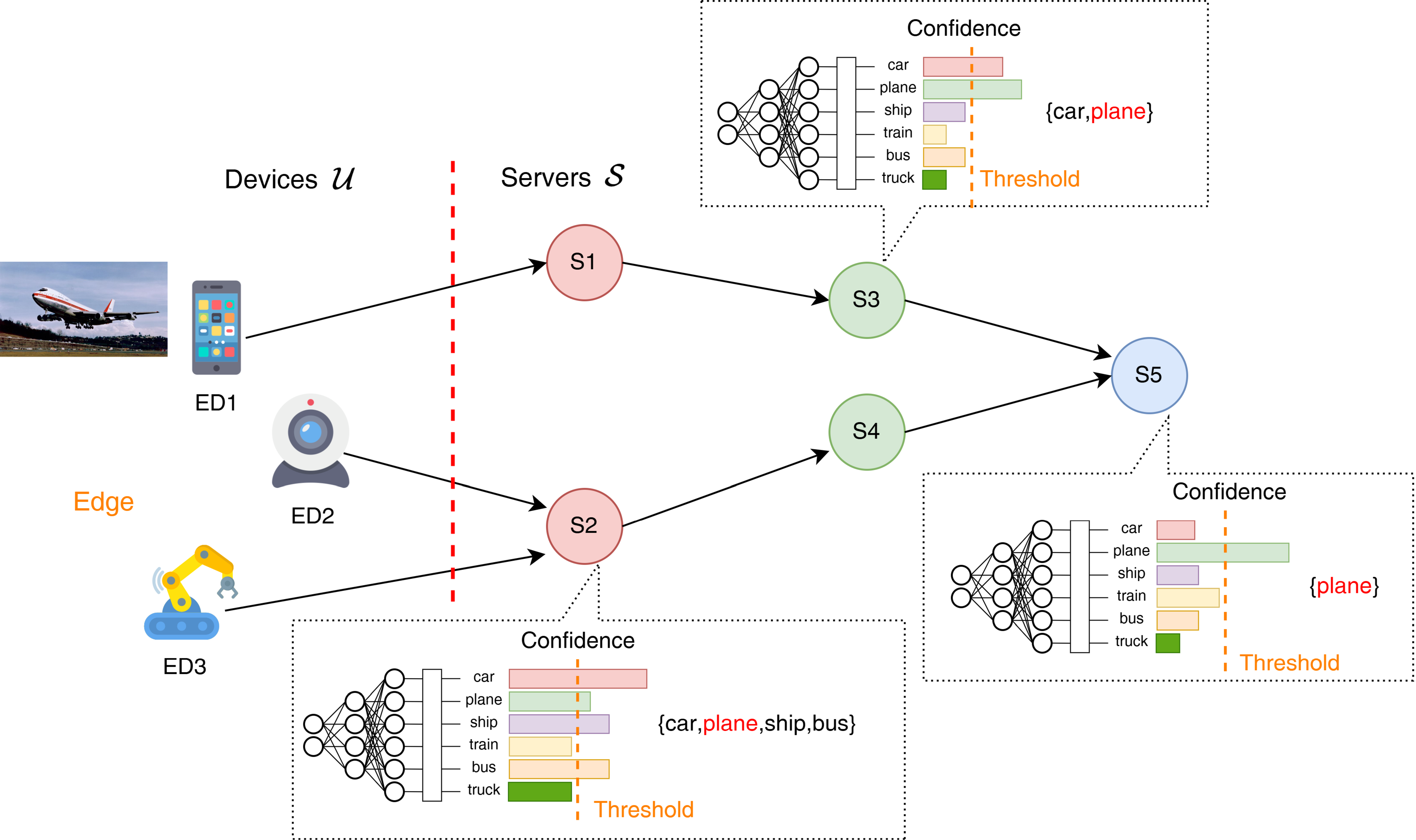}
    \caption{
    %Illustration of the class of networks under study in which 
    Edge devices task the servers at the network edge, or in the cloud, to carry out some inference task. Cloud servers typically entails larger latency and energy consumption but, potentially, also a better inference.}
    \label{fig:network_model}
\end{figure*}

In this context, it is useful to revisit existing resource allocation paradigms to assess their capability to  provide optimization strategies that efficiently and \emph{reliably} manage both transmission and computational resources \cite{barbarossa2014communicating}. The general goal is minimizing operational costs -- e.g., latency, energy consumption -- while ensuring strict compliance with all required service constraints.

A standard design methodology leverages \emph{Lyapunov optimization} (LO) \cite{neely2022stochastic},   a stochastic optimization tool based on queuing theory, which addresses dynamic resource allocation in networked systems. LO has been successfully applied in various contexts, including edge intelligence  (EI) scenarios \cite{merluzzi2020wireless,chung2024energy}. The key advantage of LO lies in its ability to design low-complexity resource allocation procedures that minimize average network costs, under long-term {average} constraints. 

However, in applications with strict reliability requirements, ensuring \emph{average} performance levels is insufficient. In fact, in such settings, the network may be required to offer strict \emph{deterministic} reliability guarantees that hold even under \emph{worst-case} conditions. For example, in an autonomous driving application, it may be not enough to ensure that, on average, an image classifier returns accurate predictions of street signs. Rather, it is important that the classifier outputs reliable decisions in every session. In such cases, employing traditional LO frameworks may either fail to meet the required constraints or request an excessively complex optimization process \cite{neely2022stochastic}.

This paper proposes an extension of LO, named \emph{conformal Lyapunov optimization} (CLO), which  incorporates also worst-case deterministic  reliability constraints, by integrating LO with \emph{online conformal risk control} (O-CRC) \cite{angelopoulos2021gentle,feldman2023achieving,Zecchin2024}. O-CRC is a recently developed adaptive mechanism designed to control long-term reliability metrics in online learning environments \cite{feldman2023achieving}. O-CRC builds upon the conformal prediction (CP) framework \cite{shafer2008tutorial,angelopoulos2021gentle}, and it is applicable to scenarios where the AI decisions take the form of a \emph{prediction set}. This is the case not only of classification and regression problems, with point decisions augmented by error bars (see Figure 1 for an illustration), but also in tasks such as image segmentation or multi-label classification \cite{angelopoulos2021gentle}. Specific applications include question-and-answer use cases of large language models \cite{quach2023conformal,kumar2023conformal}. CLO endows LO with the capacity to offer deterministic performance guarantees, while extending O-CRC to address online optimization problems.

% which, for a given inference task, constructs prediction sets that contain the true outcome with a specified confidence level. While CP and its online counterpart \cite{gibbs2021adaptive} aim to control the miscoverage rate (i.e., the probability that a prediction set does not include the true outcome), O-CRC extends this to a broader class of risk functions. These functions are particularly relevant for set-based prediction tasks, such as controlling the False Negative Rate in 

% However, while O-CRC ensures long-term reliability, it is not inherently an online optimization tool and does not provide theoretical guarantees for optimal resource allocation. The key objective of this work is to develop a comprehensive resource allocation framework that integrates both LO and O-CRC. This framework will enable optimal resource allocation while simultaneously providing theoretical guarantees on strict long-term reliability constraints.

\subsection{Related Work}

\noindent \underline{\textit{Lyapunov optimization}}: LO has been widely applied in developing resource allocation strategies across various domains, including energy harvesting networks \cite{tapparello2014dynamic, Chengrun2018Lyapunov, Mao2015ALyapunov, qiu2018lyapunov}, vehicular networks \cite{jia2022lyapunov, Abdel2020Optimized, wang2024lyapunov}, and Industrial IoT \cite{Zhang2024ALyapunov}, among others.

Focusing on the MEC paradigm, numerous Lyapunov-based resource allocation strategies have been designed to dynamically optimize offloading decisions, aiming to strike the best trade-off between local and remote computation. Several notable examples demonstrate the use of LO for edge-assisted AI/ML tasks within the EI paradigm \cite{jia2022lyapunov, Dong2021Joint}. For instance, \cite{merluzzi2020wireless} introduces multiple resource allocation strategies for edge-assisted inference tasks, optimizing energy consumption, latency, and inference accuracy entirely through LO. The work in \cite{Samarakoon2020Distributed} extends LO-based strategies to incorporate performance constraints on higher-order statistical moments (e.g., outage probability), which are crucial for URLLC applications.

From a resource optimization perspective, LO has also been employed to support goal-oriented communications, a paradigm aimed at minimizing transmission resource usage by transmitting only the essential information required to complete an inference task \cite{chaccour2024less}. The work in \cite{di2023goal} presents a general LO framework for edge-assisted goal-oriented communications, while \cite{binucci2023multi} considers an LO-based resource allocation strategy leveraging convolutional neural networks. Furthermore, reference \cite{binucci2024enabling} explores LO-based strategies for goal-oriented neural network splitting \cite{matsubara2022split}.

LO techniques have also been employed in edge-assisted federated learning (FL) scenarios. In \cite{sun2021Dynamic,chung2024energy}, LO-based approaches are designed to minimize network energy consumption in FL applications, while \cite{su2024communication} leverages LO to optimize client selection for FL tasks.

Despite the significant contributions of these works in optimizing networked resource allocation across various domains, none of them explicitly address optimal resource allocation under strict long-term deterministic constraints.

\noindent \underline{\textit{Conformal Prediction and Conformal Risk Control}}: Recent literature has highlighted the effectiveness of CP for networking applications. In \cite{cohen2023Calibrating}, CP techniques -- both online and offline -- are applied to AI models designed to assist communication tasks, such as symbol demodulation and channel estimation, while  \cite{cohen2023guaranteed} explores the use of CP techniques for dynamic scheduling of URLLC traffic, ensuring reliability in latency-sensitive applications. In the context of spectrum access, authors in \cite{lee2024reliable} introduce a CRC approach for detecting occupied subbands in unlicensed spectrum access. Therein, O-CRC ensures reliable spectrum sensing by enforcing constraints on the false negative rate, thereby minimizing the likelihood of erroneously identifying an occupied spectrum portion as free.

For edge-inference scenarios, \cite{Zhu2024Federated} proposes a CP-based protocol to quantify uncertainty in federated inference tasks under noisy communication channels. In a related work, \cite{zhu2024conformal} presents a framework aimed at maximizing inference accuracy while satisfying long-term reliability and communication constraints in sensor networks equipped with a fusion center.

Among these works, only \cite{zhu2024conformal} considers system cost optimization, while the others focus solely on satisfying long-term constraints. However, \cite{zhu2024conformal} focuses on a specific decentralized inference setting, thus not addressing the general problem of resource allocation in multi-hop edge computing networks studied herein. Furthermore, the framework in \cite{zhu2024conformal} builds on online convex optimization, while the present contribution leverages LO for optimal resource allocation.

\subsection{Main Contributions}

This paper introduces CLO, a novel framework for optimal dynamic resource allocation that guarantees  deterministic reliability constraints on end-to-end decision processes. The main contributions are as follows:
\begin{itemize}
\item We develop CLO, a general resource allocation framework for edge intelligence in multi-hop networks (see Figure 1) that integrates LO \cite{neely2022stochastic} and O-CRC  \cite{feldman2023achieving}. CLO optimizes long-term average network costs, while satisfying long-term deterministic reliability constraints on the decisions taken by AI models throughout the network. 

%This represents a significant advancement over classical LO-based approaches \cite{binucci2022adaptive,binucci2023multi}, which focus solely on average reliability guarantees while minimizing expected costs.
\item We provide a theoretical analysis proving the effectiveness of CLO in meeting both deterministic and average long-term constraints.
\item We apply the framework to an edge-assisted inference scenario, where multiple devices perform their own  inference task (i.e, segmentation), possibly offloading computations to (edge/cloud) servers, under strict per-instance reliability constraints (see Figure 1). The simulation results show:
\begin{itemize}
\item the ability of CLO to efficiently optimize system resources while ensuring strict reliability guarantees;
\item the trade-offs between average resource optimization (granted by LO), and the satisfaction of deterministic reliability constraints (ensured by O-CRC); 
\item the impact of extra
deterministic reliability constraints on classical LO policies, on the trade-off between energy consumption and inference accuracy.
\end{itemize}
\end{itemize}

\subsection{Paper Organization}

The rest of the paper is organized as follows. Section \ref{sec:problem_definition} introduces the problem definition, considering a transmission model tailored to multi-hop networks, along with the associated data acquisition process and the key performance metrics of interest. Section \ref{sec:clo} presents the development of CLO, providing theoretical guarantees and highlighting its connections with LO  and O-CRC. In Section \ref{sec:Simulation Results}, we present simulation results for both single-hop and multi-hop network scenarios. Finally, Section \ref{sec:Conclusions} concludes the paper and outlines potential future research directions.

\section{Problem Definition}\label{sec:problem_definition}
In this paper, we address the problem of resource allocation for distributed inference in networked queueing systems under reliability constraints. 

\subsection{Network Model}

As depicted in Figure \ref{fig:network_model}, we consider a network described as a directed graph $\mathcal{G}=(\mathcal{N},\mathcal{E})$, with $\mathcal{N}$ denoting the set of the nodes and $\mathcal{E} \subseteq \{(n,m) : n,m \in \mathcal{N}, \text{with } n\neq m\}$ denoting the set of links. The set of the nodes is partitioned as
\begin{equation}
\mathcal{N}=\mathcal{U} \cup \mathcal{S},
\end{equation}
where $\mathcal{U}$ denotes the set of the edge devices (ED), or users, and  $\mathcal{S}$ denotes the set of the edge or cloud servers. We consider a remote inference setting scenario,
%ban in which 
where the EDs may decide to load the network with an inference problem, such as image classification, or question answering, under reliability constraints. 

Each server in the set $\mathcal{S}$ is equipped with an inference model, such as a deep neural network or a large language model, to produce decisions on data units (DU) generated by the EDs. Inference models can operate at different points on the trade-off curve between accuracy and computational cost. In particular, while we allow for a generic distribution of computational resources across servers, in practice servers can be organized in a hierarchical topology with more powerful servers being further from the ED (see Figure \ref{fig:network_model}) \cite{ren2019survey}, and possibly affected by a higher (transmission) latency.

\subsection{Data Acquisition and Processing}
We consider a discrete-time axis with time-slots indexed by $t=1,2,\dots$, and each time-slot characterized by a fixed duration $\delta$. For each time-slot, each $k$-th ED may generate a new inference task $\tau^k(t)$, e.g., an image to classify or a query to answer, 
independently from each other, and with a probability $\lambda^{k} \in [0,1]$.
We denote as $A^{k}(t) \in \{0,1\}$ 
%$\sim$ Bern($\lambda^k$) 
\textcolor{black}{the binary random variable indicating the arrival of a new task $\tau^k(t)$, and of the corresponding data-unit (DU) for the $k$-th device at $t$-th slot}, and we collect the arrival processes of all the users in a random vector $\mathbf{A}(t)=\{A^k(t)\}_{k=1}^{K}$. In order to forward the inference task to the network, the ED produces a DU with $W^k$ bits encoding the task $\tau^k(t)$. The tasks generated at time $t$ by all the users are collected in the vector $\mathbf{T}(t)=\{\tau^k(t)\}_{k=1}^{K}$.

The DU encoding task $\tau^k(t)$ is routed to a server $s \in \mathcal{S}$, which implements the inference task. The decision is made at some later time, described by the variable $T_{\mathrm{dec}}^{k}(t)\geq t$, after the received DU is processed by server $s$. The quality of this decision depends on the complexity of the model deployed at server $s$ and on the difficulty of the task $\tau^k(t)$. This decision quality for any inference task $\tau$ at each server $s$, is summarized by a loss function $L_s(\tau,\theta)$, which is assumed to be further controllable by a hyperparameter $\theta$.

As further detailed next, the hyperparameter $\theta$ provides a measure of the conservativeness on the decision made at the server $s$, with a smaller value of $\theta$ leading to more conservative, and thus more reliable, decisions. Mathematically, we assume that the loss function $L_s(\tau,\theta)$ is non-decreasing with respect to the hyperparameter $\theta$, and is bounded in the set $[0,1]$ (see Assumption 1 below). 

\subsection{Timeline}

% \begin{figure}
%      \centering
%      \begin{subfigure}[b]{0.5\textwidth}
%          \centering
%          \includegraphics[width=\textwidth]{Figures/GeneralFigures/SlowTrajectoryTracking.eps}
%          \caption{}
%          \label{fig:slow_track}
%      \end{subfigure}
%      \hfill
%      \begin{subfigure}[b]{0.5\textwidth}
%          \centering
%          \includegraphics[width=\textwidth]{Figures/GeneralFigures/FastTrajectoryTracking.eps}
%          \caption{}
%          \label{fig:fast_track}
%      \end{subfigure}
%         \caption{Example of a time series forecasting task where the goal is to predict the values of a continous process $y(t)$ on the basis of the past observations. The time axis is organized in frames indexed by $f=1,2,\dots$, each with a fixed duration of $S$ slots. At the end of each time frame $f$ the average tracking error $\overline{e}(f)$ is evaluated. For this specific task, this information is and then employed to plan future control actions. In case of smooth processes (Figure \ref{fig:slow_track}) the error can be averaged on longer time frames. Otherwise, in case of processes characterized by higher variability (Figure \ref{fig:fast_track}), we would consider the evaluation of the average error on shorter time frames.}
%         \label{fig:trajectory_tracking}
% \end{figure}

The time slots are partitioned in frames $f=0,1,\dots,$, each one composed of $S$ time slots. Thus, considering a time horizon of $T$ slots, we have $F=T/S$ frames. The frames act as monitoring time units  within which the network evaluates inference performance. The rationale for defining this quantity is that, for any given application, the performance of interest is the average performance across the frame. On the basis of the average performance accrued within a frame, future control actions may be planned. As an example, consider real-time visual tracking  for micro aerial vehicles \cite{li2017visual}. In this application, it is critical to monitor the average tracking error on suitably chosen time windows in order to take the control actions that are necessary to track the object of interest in future instants.

%Another example can be find in the context of moving average (MA) models \cite{shumway2000time}, whose aim is to predict the future values of a time series by averaging over the prediction errors over the previous $S$ steps. 

%The window size $S$ can be set according to the specific scenario of interest. Considering time series forecasting, if we are interested on making predictions on time series characterized by a higher variability, a shorter monitoring frame could be considered to better tracking higher variations of the process of interest. Otherwise, if the phenomena under consideration is sufficiently smooth, a longer frame can be considered.

%{\color{red} Figure 2 did not seem very useful to me}

\subsection{Reliability and Precision}\label{sec:reliability_and_precision}

To elaborate on the definition of the loss function $L_s(\tau,\theta)$, consider the image classification task depicted in Figure \ref{fig:network_model}. In this case, given an input image $x$, the goal of the server $s$ is to produce a subset $\mathcal{C}(x,\theta)$ of possible labels $y \in \mathcal{Y}$ as a function of the hyperparameter $\theta$. For instance, following the conformal prediction (CP)\cite{angelopoulos2021gentle} framework \cite{gibbs2021adaptive}, the hyperparameter $\theta$ represents a threshold on the confidence level produced by the inference model, and the prediction set is given by
\begin{equation}
\label{eq:p_set_image_classification}
    \mathcal{C}(x,\theta)=\{y \in \mathcal{Y} : p(y|x)\geq\theta\},
\end{equation}
with $p(y|x)$ denoting the confidence level associated by the inference model to the label $y$, taking values in the set $\mathcal{Y}$ for input $x$. In this case, the loss function is typically given as the miscoverage loss
\begin{equation}
\label{eq:acp_loss}
    L_s(x,\theta)=\mathbbm{1}(y_{\mathrm{true}} \notin\mathcal{C}(x,\theta)),
\end{equation}
where $y_{\mathrm{true}}$ is the true label associated to the input, and $\mathbbm{1}\{\cdot\}$ is the indicator function, which equals to $1$ if the argument is true and $0$ otherwise. By \eqref{eq:p_set_image_classification}, the loss \eqref{eq:acp_loss} increases with the hyperparameter $\theta$, as required.

As another example, take an image segmentation task for an autonomous driving scenario \cite{feldman2023achieving}. In this application, given an input image $x$, the prediction is given by a binary mask identifying the pixels of the image belonging to obstacles.
This decision is typically obtained as
\begin{equation}
\label{eq:p_set_seg}
    \mathcal{C}(x,\theta)=\{(i,j) : p(i,j|x)\geq \theta\},
\end{equation}
 where $(i,j)$ are the pixels coordinates, and $p(i,j|x)$ is the estimated probability that pixel $(i,j)$ belongs to an obstacle \cite{angelopoulos2022conformal}. 
 In this case, the loss is typically given by the false negative rate (FNR), given by the fraction of pixels belonging to the obstacle that are not included in the set $\mathcal{C}(x,\theta)$, i.e., 
 \begin{equation}
 \label{eq:fnr_loss}
     L_{s}(x,\theta)=\frac{\abs{y_{\mathrm{true}} \cap \mathcal{\overline{C}}(x,\theta)}}{\abs{y_{\mathrm{true}}}},
 \end{equation}
where $y_{\mathrm{true}}$ is the set of pixels including the object of interest and $\mathcal{\overline{C}}(x,\theta)$ is the complement of set $\mathcal{C}(x,\theta)$.
 The FNR \eqref{eq:fnr_loss} is also an increasing function of the hyperparameter $\theta$.
% \begin{equation}
% \label{eq:loss_seg}
%     L_s(x,\theta)=1-\frac{\abs{\mathcal{C}(x,\theta)\cap y_{t}}}{\abs{y_{t}}},
% \end{equation}

By the mentioned monotonicity assumption on the loss $L_s(\tau,\theta)$, a higher reliability (e.g., a lower loss) can be guaranteed by reducing the hyperparameter $\theta$. Specifically, we make the following assumption, which is satisfied in the two examples discussed above.

\begin{assumption} The reliability loss function $L_s(\tau,\theta)$ is non-decreasing in the hyperparameter $\theta$ for each server $s \in \mathcal{S}$ and for each task $\tau$. Furthermore, it is bounded in the interval $[0,1]$, and it satisfies the equality
\begin{equation}\label{ass:reliability_assumption}
    L_s(\tau,0)=0, \hspace{6pt} \mathrm{\textnormal{for each }} s \in \mathcal{S} \textnormal{ and } \tau.
\end{equation}
\end{assumption}

While increasing reliability, a smaller hyperparameter $\theta$ yields a less informative, or precise, decision. For example, in image classification and segmentation, a small $\theta$ entails larger prediction sets \eqref{eq:p_set_image_classification} and \eqref{eq:p_set_seg}. Accordingly, there is a trade-off between reliability (e.g., true pixels in the prediction set) and precision (e.g., correct pixels w.r.t. the set cardinality). 

To capture this trade-off, we introduce the precision loss $F_s(\tau,\theta)$, which satisfies the following assumption.

\begin{assumption}\label{ass:precision_assumption} The precision loss function $F_s(\tau,\theta)$ is non-increasing in the hyperparameter $\theta$ for each $s \in \mathcal{S}$ and for each task $\tau$. Furthermore, it is bounded in the interval $[0,1]$, and it satisfies the equality
\begin{equation}
    F_s(\tau,0)=1, \hspace{6pt} \textnormal{for each } s \in \mathcal{S} \textnormal{ and } \tau.
\end{equation}
\end{assumption}

For example, for classification tasks, one can adopt the precision loss
\begin{equation}
    \label{eq:imprecision_definition}
    F_{s}(x,\theta)=\frac{\abs{\mathcal{C}(x,\theta)}}{\abs{\mathcal{Y}}},
\end{equation}
where $\abs{\mathcal{Y}}$ is the size of the output space $\mathcal{Y}$, while $\abs{\mathcal{C}(x,\theta)}$ the size of the prediction set \eqref{eq:p_set_image_classification}. For image segmentation, a widely used precision loss is the false positive rate (FPR)
\begin{equation}\label{eq:precision}
F_s(x,\theta)=\frac{\abs{\overline{y}_{\mathrm{true}}\cap\mathcal{C}(x,\theta)}}{\abs{\overline{y}_{\mathrm{true}}}},
\end{equation}
i.e., the fraction of pixels of the estimated target that are outside the true target, e.g., in the set $\overline{y}_{\mathrm{true}}=\mathcal{Y}\setminus y_{\mathrm{true}}$.

\textcolor{black}{Appendix \ref{sec:mon_proofs} reports the proofs of monotonicity for the presented precision and reliability losses.}.

\subsection{Transmission Model}

\begin{figure*}[ht]
    \centering
    \includegraphics[width=0.75\linewidth]{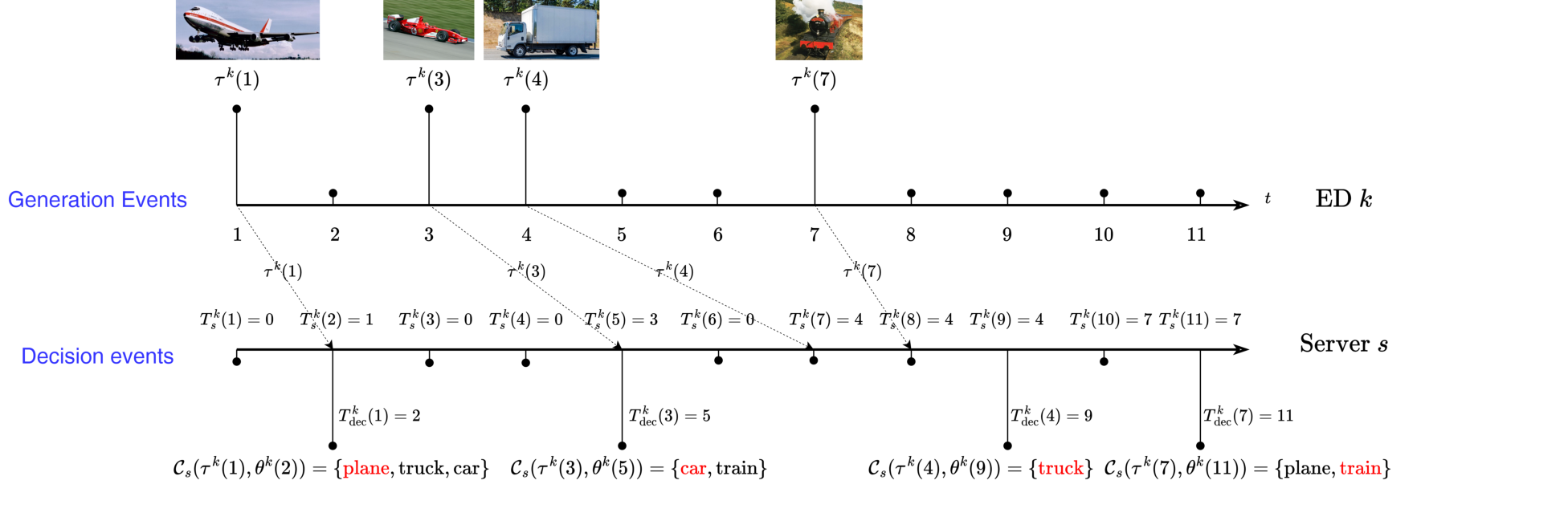}
    \caption{Sequence of DUs generated at $k$-th ED and associated decisions at server $s$ (ignoring other EDs and servers). 
    %The time-axis at the top shows the generation process of new DUs at the ED $k$. The time-axis at the bottom shows the temporal evolution of the decision process at the server $s$. 
    At any time $t$, the server may decide on the DU at the head of its queue $Q_{s}^{k}(t)$, whose generation time is encoded by  $T_{s}^{k}(t)$.}
    \label{fig:temporal_diagram}
\end{figure*}

The transmission phase follows a standard queuing model for multi-hop wireless networks \cite{neely2022stochastic}. In each slot $t$, the link $(n,m) \in \mathcal{E}$ is described by the channel state $S_{n,m}(t)$, and the overall state matrix is $\mathbf{S}(t)=\{S_{n,m}(t)\}_{(n,m)\in\mathcal{E}}$. A power allocation matrix $\mathbf{P}(t)=\{P_{n,m}(t)\}_{(n,m)\in \mathcal{E}}$ determines the power $P_{n,m}(t)$ allocated on each edge $(n,m)$ at time $t$. The overall power consumption of the $n$-th node in the network is given by the sum
\begin{equation}
    P_{n}(t)=\sum_{(n,m) \in \mathcal{E}}P_{n,m}(t),
\end{equation}
which must satisfy the constraint $P_{n}(t)\leq P_{n}^{\max}$.

Given the allocated powers $\mathbf{P}(t)$ and states $\mathbf{S}(t)$, the transmission rate on each link $(n,m) \in \mathcal{E}$ at time $t$ is given by
\begin{equation}
    \mu_{n,m}(t)=C_{n,m}(\mathbf{P}(t),\mathbf{S}(t)),
\end{equation}
for some capacity function $C_{n,m}(\cdot)$. For example, in AWGN channels without interference, according to Shannon theory the capacity function can be chosen as \cite{shannon1948mathematical}
\begin{equation}
    C_{n,m}(t)=B_{n,m}\log_{2}\left(1+\frac{P_{n,m}(t)S_{n,m}(t)}{B_{n,m}N_0}\right),
\end{equation}
where $B_{n,m}$ represents the transmission bandwidth for the link $(n,m)$, while $N_0$ is the noise power spectral density.

Recalling that $W^k$ represents the size in bits of the DUs generated by the $k$-th user, the transmission delay of a DU generated by the $k$-th user across the link $(n,m)$ is given by
\begin{equation}
\label{eq:tranmission_time_inequality}
    D_{n,m}^{k}(t)=\frac{W^{k}}{C_{n,m}(t)},
\end{equation}
which we assume to be no longer than the duration $\delta$ of the time slot. Thus, the energy required to forward a DU of the $k$-th ED at the $t$-th slot is expressed by 
\begin{equation}
    E_{n,m}^{k}(t)=P_{n,m}(t)D_{n,m}^{k}(t).
\end{equation}

Indicating with $R_{n,m}^{k}(t)$ the binary variable capturing if the link $(n,m)$ is used for the transmission of a DU by the $k$-th ED in the time slot $t$, i.e., 
\begin{equation}
    R_{n,m}^{k}(t)=\begin{cases}
			1, & \text{link $(n,m)$ carries a DU of the $k$-th ED}\\
            0, & \text{otherwise},
		 \end{cases}
   \label{eq:limited_link}
\end{equation}
we can constraint the maximum number of DUs that can be sent on any link $(n,m)$, by
\begin{equation}
    \sum_{k=1}^{K}R_{n,m}^{k}(t)\leq R_{n,m}^{\max}\hspace{12pt} \forall \; (n,m), t. \hspace{22pt}.
\end{equation}
% We further impose the following constraint on the transmission rate
% \begin{equation}
% \label{eq:tx_rate_constraint}
% \mu_{n,m}(t)\geq \sum_{k=1}^K \frac{W^k}{\delta}R_{n,m}^k(t),
% \end{equation}
% which means that, if $R_{n,m}^{k}(t)=1$ for a particular $k$, we need to guarantee a rate which ensures the transmission of a DU of $W^k$ bits across the link $(n,m)$ within a time-slot.  
The overall energy consumed throughout the network at the $t$-th time-slot is given by
\begin{equation}
\label{eq:overall_tx_energy}
    E_{\mathrm{tot}}(t)=\sum_{k=1}^{K}\sum_{(n,m)\in \mathcal{E}}R_{n,m}^{k}(t)E_{n,m}^{k}(t).
\end{equation}
\subsection{Edge Inference and Queueing Model}
At any time-slot, each server $s$ decides to process a number of DUs in its queues, along with the corresponding inference tasks. To describe this decision, we introduce the binary variable 
\begin{equation}
    I_{s}^{k}(t)=\begin{cases}
			1, & \text{if server $s$ processes a task for the $k$-th ED}\\
            0, & \text{otherwise}.
		 \end{cases}
\end{equation}
We impose that, at each time slot, each server $s$ can process at most $I_s^{\max}$ tasks, i.e., 
\begin{equation}
    \sum_{k=1}^{K}I_{s}^{k}(t)\leq I_s^{\max} \hspace{6pt} \forall \; s, t.
\end{equation}

We assume that the DUs injected by the EDs into the network are buffered into separate transmission queues. Specifically, the $n$-th node has a dedicated queue $Q_{n}^{k}(t)$ for the traffic of the  $k$-th ED, which reflects the number of queued DUs. Note that an ED can also potentially serve, as an intermediate node, for the traffic of other EDs. 
%Denote as $Q_{n}^{k}(t)$ the state of the queue of the $n$-th node associated to the $k$-th ED, which is measured in number of DUs. 

The evolution of each queue is given by 
\begin{equation}
\begin{split}
    Q_{n}^{k}(t+1)&=\max\left(0,Q_{n}^{k}(t)-\hspace{-10pt}\sum_{(n,m)\in\mathcal{E}}\hspace{-10pt}R_{n,m}^{k}(t)-\mathbbm{1}\{n \in \mathcal{S}\}I_{n}^{k}(t)\right)\\
    &+A^{n}(t)\mathbbm{1}\{n \in \mathcal{U}\}+\sum_{(l,n) \in \mathcal{E}}\hspace{-5pt}R_{l,n}^{k}(t).
    \end{split}
    \label{eq:system_queues}
\end{equation}
For each time slot $t$, the queue is updated by subtracting the number of outgoing DUs, given by $\sum_{(n,m)\in\mathcal{E}} R_{n,m}^{k}(t)$, and, if node $n$ is a server (i.e., $n \in \mathcal{S}$), by the number of processed DUs, $I_{n}^{s}(t)$. Conversely, it is incremented by the number of task arrivals at the ED, if $n \in \mathcal{U}$, and by the DUs received from other nodes. Since a DU can be processed only if the corresponding queue is not empty, we have the implication
% The first term in \eqref{eq:system_queues} accounts for the number $\sum_{(n,m)\in\mathcal{E}}R_{n,m}^{k}(t)$ of outgoing DUs and for the number $I_{n}^{s}(t)$ of processed DUs if the node is a server, i.e., $n \in \mathcal{S}$. The second term in \eqref{eq:system_queues} quantifies the number of tasks arrivals at the ED, if  $n \in \mathcal{U}$, and the third term corresponds to traffic incoming from other nodes. Since a DU can be processed only if the corresponding queue is not empty, we have the implication
\begin{equation}
 Q_s^k(t)=0 \implies I_s^k(t)=0.   
\end{equation}
In a similar way, we also have
\begin{equation}
    Q_n^k(t)=0 \implies R_{n,m}^k(t)=0 \hspace{5pt}\forall m  \hspace{2pt} : \hspace{2pt}  (n,m) \in \mathcal{E},
\end{equation}
since no DU can be sent to an outgoing link if the corresponding queue is empty.

In the setting under study, it is important to keep track not only of the number of DUs in the queues via \eqref{eq:system_queues}, but also of their identities. To this end, we define the variable $T_{s}^k(t)$ as the generation time of the DU at the head of the queue of the $s$-th server, associated to the $k$-th ED, at time $t$. When the queue is empty we simply set $T_{s}^k(t)=0$. Figure \ref{fig:temporal_diagram} illustrates the temporal evolutions of the DUs possibly generated at the $k$-th ED, as well as the corresponding timings of the decisions at the $s$-th server. Note that, for simplicity, the figure considers a simplified situation in which all the DUs of the $k$-th ED are processed by the same server $s$, which is not the general case.

\subsection{Performance Metrics}
The design goal is to minimize a weighted objective encompassing the transmission energy \eqref{eq:overall_tx_energy} and the overall precision loss, under strict reliability constraints. To this end, we optimize over the sequence of transmission scheduling $\mathbf{R}(t)=\{R_{n,m}^{k}(t)\}_{(n,m) \in \mathcal{E}, k \in \mathcal{U}}$, the transmission powers $\mathbf{P}(t)=\{P_{n,m}(t)\}_{(n,m)\in\mathcal{E}}$, and the task assignments $\mathbf{I}(t)=\{I_{s}^{k}(t)\}_{s \in \mathcal{S}, k \in \mathcal{U}}$. As detailed below, we also introduce a sequence of variables $\mathbf{\Theta}(t)=\{\theta^k(t)\}_{k=1}^{K}$, one for each ED, that, according to Section \ref{sec:reliability_and_precision}, are used to define the level of conservativeness applied by the server $s$ when it processes tasks for the $k$-th ED.

We impose the deterministic worst-case constraint that, as time goes on, the average reliability loss in each frame for the decisions made on tasks belonging to the $k$-th ED is increasingly closer to a target value $r^k$.  Mathematically, this requirement is formulated as
\begin{equation}
\label{eq:reliability_const}
    \overline{L}^k=\frac{1}{F}\sum_{f=0}^{F-1}\frac{1}{N_f^k}\sum_{t=fS+1}^{(f+1)S}\sum_{s \in \mathcal{S}}I_{s}^{k}(t)L_{s}^{k}(t) \leq r^k+\mathcal{O}\left(\frac{1}{F}\right),
\end{equation}
where 
\begin{equation}
\label{eq:loss_per_dev}
    L_{s}^{k}(t)=L_{s}(\tau^k(T_{s}^k(t)),\theta^k(t))
\end{equation} 
is the loss accrued by a decision taken at time $t$ by the server $s$ on the task  $\tau^k(T_s^k(t))$; the function $\mathcal{O}(\frac{1}{F})$ tends to zero as $F\to\infty$; and the quantity
\begin{equation}
N_f^k=\displaystyle\sum_{t=fS+1}^{(f+1)S}\sum_{s \in \mathcal{S}}I_{s}^{k}(t) 
\end{equation}
 denotes the number of DUs of the $k$-th ED, whose decisions on have been taken within the $f$-th frame. Importantly, the constraint defined in \eqref{eq:reliability_const} must be satisfied deterministically for each run of the optimization protocol. To this end, the network controls the risk tolerance of the decisions made for each ED $k$ via the sequence of variables $\theta^k(t)$. 

The optimization objective is given by the weighted sum of the transmission energy \eqref{eq:overall_tx_energy} and of the overall precision loss across all the EDs, i.e.,  
\begin{equation}
    J(t)=E_{\mathrm{tot}}(t)+\eta F_{\mathrm{tot}}(t),
    \label{eq:cost_function}
\end{equation}
where $\eta\geq0$ is a multiplier used to explore the energy/precision trade-off. The overall precision loss is given by 
\begin{equation}
    \label{eq:objective_function}
    F_{\mathrm{tot}}(t)=\sum_{k=1}^{K}\sum_{s \in \mathcal{S}} I_{s}^{k}(t)F_{s}^{k}(t),
\end{equation}
with
\begin{equation}
\label{eq:imprecision_per_dev}
    F_{s}^k(t)=F_{s}(\tau^k(T_{s}^{k}(t)),\theta^k(t))
\end{equation}
denoting the precision loss accrued by the decision taken by the server $s$ on the DU $\tau^k(T_{s}^{k}(t))$.

\subsection{Problem Formulation}

Overall, we aim to addressing the optimization problem
\begin{mini}
    {{\mathbf{\Phi}(t)}}{\lim_{T\to\infty}\frac{1}{T}\sum_{t=1}^{T}\mathbb{E}\{J(t)\}}{\label{eq:long_term_prob}}{}
    \addConstraint{\text{(a)}\;}{\text{long-term reliability constraints \eqref{eq:reliability_const} }\; \forall k}{}{}
    \addConstraint{\text{(b)}\;}{Q_{n}^{k}(t) \text{ are mean-rate stable } \hspace{6pt} \forall k,n \footnotemark}{}
        \addConstraint{\text{(c)}\;}{P_{n}(t) \leq P_{n}^{\max}\hspace{27pt} \forall n,t}{}
    \addConstraint{\text{(d)}\;}{\sum_{k=1}^{K}I_{s}^{k}(t) \leq I_s^{\max} \hspace{16pt} \forall s,t} 
    \addConstraint{\text{(e)}\;}{\sum_{k=1}^{K}R_{n,m}^{k}(t)\leq R_{n,m}^{\mathrm{max}} \hspace{22pt} \forall (n,m) \in \mathcal{E},t}{}
\end{mini}
where $\mathbf{\Phi}(t)=\{\mathbf{I}(t),\mathbf{R}(t),\mathbf{P}(t),\mathbf{\Theta}(t)\}$ denotes the set of the optimization variables. Via problem \eqref{eq:long_term_prob}, we aim to minimize the average energy/precision trade-off $J(t)$ under (a) long-term deterministic reliability constraints; (b) mean-rate stability of all the queues; (c) transmission power constraint for each device; (d) maximum processing capabilities for each server; (e) maximum transmission capacity for each link.

\footnotetext{Mean-rate stability is a standard requirement in stochastic optimization of networked queuing systems \cite{neely2022stochastic}.}

The goal is to solve problem \eqref{eq:long_term_prob} through an online optimization strategy, which is adaptive with respect to the dynamics of the system. To this end, at every time instant $t$, a central controller observes the system state, defined by the state of all the queues and channels, and chooses the control variables $\mathbf{\Phi}(t)$. Distributed implementations are also possible, and are left for future investigations.

\section{Conformal Lyapunov Optimization}\label{sec:clo}

In this section, we describe and analyze the proposed CLO algorithm, which addresses problem \eqref{eq:long_term_prob} by integrating LO \cite{neely2022stochastic} and O-CRC \cite{feldman2023achieving}.

\subsection{An Overview of Conformal Lyapunov Optimization}

\begin{figure*}[ht]
    \centering
    \includegraphics[width=0.55\linewidth]{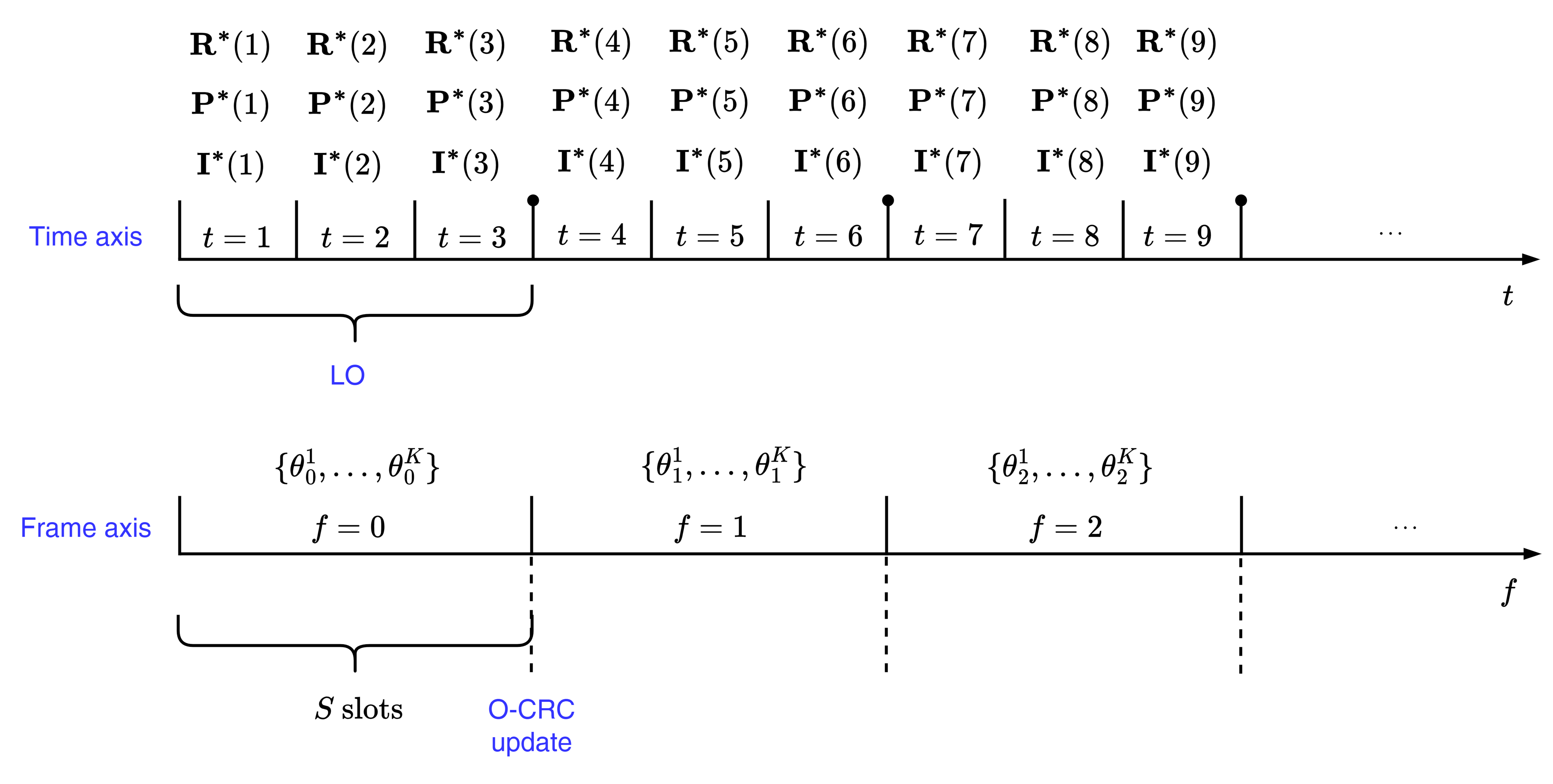}
    \caption{CLO frame-based structure: each frame $f\in \mathbb{N}_0$ is composed by $S$ slots, with fixed duration. In each time slot $t$ within the $f$-th frame, the powers $\mathbf{P}(t)$, the scheduling $\mathbf{R}(t)$, and server assignments $\mathbf{I}(t)$ are obtained by LO for fixed reliability parameters $\mathbf{\Theta}_f=\{\theta^k_f\}_{k=1}^{K}$, which are updated at the end of each frame by O-CRC, to address the reliability constraint \eqref{eq:reliability_const} .}
    \label{fig:time-axis-organization}
\end{figure*}

%The long-term reliability constraint (\ref{eq:long_term_prob}a) cannot be addressed within the classical LO framework. In fact, 
Classical LO only supports \emph{statistically}-average long-term constraints, while it cannot address \emph{deterministic} (worst-case) long-term reliability constraints of the form (\ref{eq:long_term_prob}a). Conversely, O-CRC targets deterministic constraints as in (\ref{eq:long_term_prob}a), but it is not designed to tackle optimization problems, focusing instead only on inference reliability. \textcolor{black}{The key difference between a statistical average constraint and a deterministic reliability constraint is that  the former is satisfied on average across multiple runs of the optimization procedure. Therefore, it is generally violated in any specific run of the system. In contrast, a deterministic reliability constraint is more stringent, as it demands that the given reliability condition be satisfied for each individual run of the procedure.
}

A key observation is that, if we removed the constraint (\ref{eq:long_term_prob}a) from problem \eqref{eq:long_term_prob} and we fixed the reliability-controlling variables $\mathbf{\Theta}(t)$, LO would be directly applicable as a solution method to optimize over the remaining variables $\{\mathbf{P}(t),\mathbf{I}(t),\mathbf{R}(t)\}$. Based on this observation, CLO tackles the problem \eqref{eq:long_term_prob} including the constraint (\ref{eq:long_term_prob}a) by applying LO within each frame assuming fixed reliability variables, and then updating the reliability variables at the end of the frame employing a rule inspired by O-CRC. 

As shown in Figure \ref{fig:time-axis-organization}, the reliability variables $\{\theta^k_f=\theta(fS+1)\}_{k=1}^{K}$ are fixed at the beginning of a frame, and LO is applied to address problem \eqref{eq:long_term_prob} without the reliability constraint (\ref{eq:long_term_prob}a). In order to meet the long-term reliability constraint (\ref{eq:long_term_prob}a), the variables $\{\theta^k_f\}_{k=1}^{K}$ are then updated at the end of the frame by using feedback about the decisions made within the frame. Intuitively, the updates should decrease the variables $\theta^k_f$ if the decisions for the $k$-th ED have been too inaccurate during the $f$-th frame, requiring an increase of the conservativeness for the inference outputs of the $k$-th ED's tasks. 

The next subsections will provide deeper insights on the CLO algorithm, which is also detailed in Algorithm \ref{alg:clo}. We start outlining how to update the reliability variables across frames according to O-CRC; and then showing how to adapt the LO framework for optimal power, transmission scheduling, and inference allocations, within each frame. Finally, we provide a theoretical analysis that proves the effectiveness of the proposed approach. 

\subsection{Updating Reliability Parameters}\label{sec:ocrc_update}

As overviewed in the previous subsection, the proposed CLO updates the variables $\mathbf{\Theta}_f=\{\theta^k_f\}_{k=1}^{K}$ at the end of each frame $f \in\mathbb{N}_0$ to address reliability constraints (\ref{eq:long_term_prob}a). 
% To elaborate, write as $N_f^k$ the number of DUs generated by the ED $k$ that are classified within the frame $f$. 
CLO assumes the availability of feedback about the average loss accrued by these decisions \textcolor{black}{with a delay of $d^k$ time-slots for each device $k$. This delay may result from the overhead associated with estimating and disseminating frame loss information. Accordingly, the update of reliability hyperparameters at frame \( f \) is based on the average reliability loss observed at frame \( f - d^k \) for each device $k$.}

The average loss  at frame $f$ is obtained by summing the losses $L_s^k(t)$ in \eqref{eq:loss_per_dev} for all DUs processed within the slots of the $f$-th frame, i.e., 
\begin{equation}
    \label{eq:feedback}
    \overline{L}_f^k=\frac{1}{N_f^k}\sum_{t=fS+1}^{(f+1)S}\sum_{s \in \mathcal{S}}I_s^k(t)L_s^k(t).
\end{equation}
In practice, the feedback \eqref{eq:feedback} may be obtained by recording the outcomes of the inference decisions. 
For instance, for the inference task of predicting the trajectory of an object in motion, the subsequent observation of the object's movement can confirm whether the object pixels are included or not in the decision set, yielding the loss $L_s^k(t)$ \cite{dixit2023adaptive}.

Based on the received feedback at frame $f$, CLO updates the reliability variables as \cite{feldman2023achieving}
\begin{equation}
\label{eq:hyperparameters_update}
\textcolor{black}{\theta^k_{f+1}=\theta^k_{f}+\gamma^k\mathbbm{1}\{N^k_{f-d^k}>0\}(r^k-    \overline{L}^k_{f-d^k}),}
\end{equation}
where $\gamma^k>0$ is the learning rate. By \eqref{eq:hyperparameters_update}, if the reliability constraint $r^k$ is violated within the $(f-d^k)$-th frame, i.e., if $\overline{L}^k_{f-d^k}>r^k$, the variable $\theta^k_f$ is decreased, i.e., $\theta^k_{f+1}\leq\theta^k_f$. This leads to more conservative, and thus less precise, decisions for the $k$-th ED during the next $(f+1)$-th frame. Conversely, when the reliability constraint is satisfied within the $(f-d^k)$-th frame, i.e., $\overline{L}^k_{f-d^k}<r^k$, the parameter $\theta^k_f$ is increased by the update \eqref{eq:hyperparameters_update}, prioritizing precision over reliability.

An important remark pertains the impact of the frame size $S$ on the update \eqref{eq:hyperparameters_update}. Indeed, larger frame sizes $S$ entails a more informative feedback \eqref{eq:feedback}, since the loss is averaged over a larger number of decisions. On the other hand, having larger frames, thus a less frequent update of $\theta^k_{f}$, will proportionally increase the overall number of time slots before the updates \eqref{eq:hyperparameters_update} will converge to a stable solution, satisfying the reliability constraint (\ref{eq:long_term_prob}a).

The resulting tension between informativeness of each update and update rate (i.e., convergence rate) will be studied theoretically in Section \ref{sec:theoretical_guarantees}.

\subsection{Within-Frame Optimization of  Power Allocation and Transmission/ Inference Scheduling}

We now focus on the optimal power allocation and optimal transmission/inference scheduling within each frame $f$. To this end, CLO addresses problem \eqref{eq:long_term_prob} without the reliability constraint (a), while fixing the reliability variables $\mathbf{\Theta}_f$. This problem is tackled via LO, which solves a static problem at each time slot $t$ over the optimization variables $\{\mathbf{I}(t),\mathbf{R}(t),\mathbf{P}(t)\}$.

Specifically, at each time $t$, LO addresses the instantaneous problem
\footnote{Derivations are detailed in Section I of the supplemental material.}
\begin{equation}
    \begin{split}
        \min_{\{\mathbf{P}(t),\mathbf{I}(t),\mathbf{R}(t)\}}\; &VJ(t)-\hspace{-15pt}\sum_{(n,m) \in \mathcal{E}, k \in \mathcal{U}}\hspace{-15pt}U_{n,m}^{k}(t)R_{n,m}^{k}(t)\\
        &-\hspace{-12pt}\sum_{n \in \mathcal{N},k\in \mathcal{U}}\hspace{-10pt}\mathbbm{1}\{n \in \mathcal{S}\}Q_{n}^k(t)I_n^k(t)\\
        & \text{s.t.} \quad \text{(\ref{eq:long_term_prob}c)-(\ref{eq:long_term_prob}e)},
    \end{split}
    \label{eq:upper_bound}
\end{equation}
% \begin{equation}
%     \begin{split}
%         \min_{\{\mathbf{P}(t),\mathbf{I}(t),\mathbf{R}(t)\}}\; &V\eta\hspace{-5pt}\sum_{s \in \mathcal{S}, k \in \mathcal{U}}\hspace{-6pt}I_{s}^{k}(t)F_{s}^{k}(t)+V\hspace{-15pt}\sum_{(n,m)\in\mathcal{E}, k \in \mathcal{U}}\hspace{-15pt}R_{n,m}^{k}(t)E_{n,m}(t)\\
%         &-\hspace{-15pt}\sum_{(n,m) \in \mathcal{E}, k \in \mathcal{U}}\hspace{-15pt}U_{n,m}^{k}(t)R_{n,m}^{k}(t)-\hspace{-12pt}\sum_{n \in \mathcal{N},k\in \mathcal{U}}\hspace{-10pt}\mathbbm{1}\{n \in \mathcal{S}\}Q_{n}^k(t)I_n^k(t)\\
%         & \text{s.t.} \quad \text{(\ref{eq:long_term_prob}c) - (\ref{eq:long_term_prob}e)},
%     \end{split}
%     \label{eq:upper_bound}
% \end{equation}
where $V>0$ is a hyperparameter that trades energy consumption and precision, for queues congestion (average delay), and
\begin{equation}
    U_{n,m}^{k}(t)=Q_{n}^{k}(t)-Q_{m}^{k}(t)
\end{equation}
is the differential backlog on link $(n,m)\in\mathcal{E}$ for ED $k$.

The objective function in \eqref{eq:upper_bound} is a weighted sum of the current contribution $F_{\mathrm{tot}}(t)$ to the objective function in the original problem \eqref{eq:long_term_prob}, and two penalty terms. The first term  $\sum_{(n,m) \in \mathcal{E}, k \in \mathcal{U}}U_{n,m}^{k}(t)R_{n,m}^{k}(t)$, favors transmission for traffic with largest differential backlog \cite{neelydynamic}. The second term $\sum_{n \in \mathcal{N},k\in \mathcal{U}}\mathbbm{1}\{n \in \mathcal{S}\}Q_{n}^k(t)I_n^k(t)$ favors processing for the servers with the largest number of queued DUs. 

 Since the variables $\mathbf{I}(t)$ and $\mathbf{R}(t)$ take values in discrete sets, the problem \eqref{eq:upper_bound} is a mixed-integer program. Furthermore, it is convex with respect to the transmission powers $\mathbf{P}(t)$ when $\{\mathbf{R}(t),\mathbf{I}(t)\}$ are fixed. Approximation techniques, such as branch-and-bound or convex relaxation, support the evaluation of a near-optimal solution.

 \subsection{Modeling the Precision Loss}\label{subsec:model}

\begin{figure}[ht]
    \centering
    \includegraphics[width=0.90\linewidth]{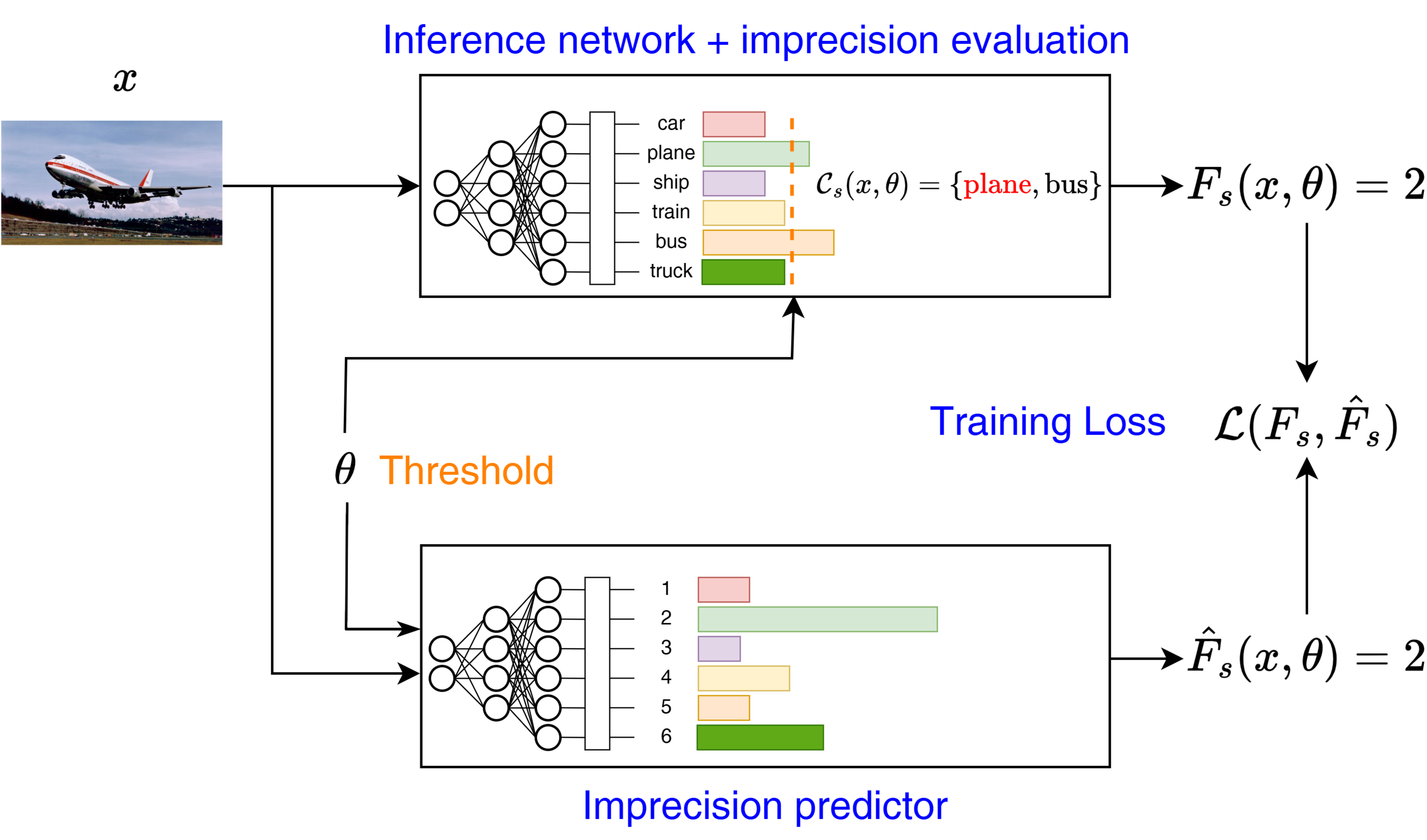}
    \caption{ Training of the NN predictor of the precision loss function ${F}_s(x,\theta)$ associated with the classifier employed at the $s$-th server, for an image classification task. The classifier input $x$ and threshold $\theta$ are the input  pair $(x,\theta)$ for the predictor, which produces an estimate $\hat{F}_s(x,\theta)$ of the precision loss value  possibly associated to the classifier decision. The training loss $\mathcal{L}(F_s,\hat{F}_s)$ evaluates the mismatch between the actual and the predicted precision loss.}
    \label{fig:imprecision_predictor_training}
\end{figure}

By their definitions in \eqref{eq:imprecision_definition}, \eqref{eq:acp_loss} and \eqref{eq:fnr_loss}, both the reliability and the precision losses associated with an inference task $\tau$, can be evaluated only after the execution of the task. This is not an issue for the reliability loss function $L_s(\tau,\theta)$. In fact, the O-CRC update \eqref{eq:hyperparameters_update} only requires feedback after a decision is implemented. In contrast, the precision loss is requested to solve the instantaneous problem  \eqref{eq:upper_bound}, which has to provide the decision variables $I_s^k(t)$ over the time slots $t$. In practice, this requires an estimate of the precision loss function before processing the inference task. 

To tackle this issue, as illustrated in Figure \ref{fig:imprecision_predictor_training}, we propose to train $|\mathcal{S}|$ neural networks (NNs) devoted to predict the precision loss associated to a pair $(\tau,\theta)$ for each of the $|\mathcal{S}|$ servers. Specifically, the $s$-th NN predictor is associated with the inference model employed by the $s$-th server. The trained predictors $\{\hat{F}_s(\tau,\theta)\}_{s=1}^{\vert \mathcal{S} \vert}$ act as approximators of the actual precision losses \eqref{eq:imprecision_per_dev}, and they can be employed to evaluate the cost function of the instantaneous optimization problem \eqref{eq:upper_bound}. As depicted in Figure \ref{fig:imprecision_predictor_training}, a possible approach consists in training the precision loss predictor on an augmented training set of the original inference task, where we consider a set of possible values for the reliability variable $\theta$ for each training sample $\tau$. The output variable is represented by the precision loss accrued by each training pair $(\tau,\theta)$ by the actual $s$-th inference model. 

An alternative approach involves training a set of low-complexity networks through knowledge-distillation techniques \cite{gou2021knowledge}. In this setup, the actual inference models at the servers, play the role of teacher networks, while precision loss approximators act as student networks. The student models are trained to mimic the outputs of the inference models, thus allowing to obtain a reliable estimate of the effective loss. For example, in the context of prediction-set construction for image classification, a practical measure of imprecision can be obtained by counting the number of classes for which the student model assigns a confidence level exceeding a predefined threshold (see Figure \ref{fig:imprecision_predictor_training}).

\textcolor{black}{We note that alternative approximation techniques can also be considered, each with a distinct impact on the algorithm's performance. The effect of neural network based precision loss approximation is evaluated in Appendix~\ref{sec:additional_results}.
}

\begin{algorithm}[ht]
\label{alg:lyapunov_driven_rrc}
\begin{algorithmic}[1]

\STATEnonum \textbf{Input: }Graph $\mathcal{G}=(\mathcal{N},\mathcal{E})$; time frame duration $S$; and step-sizes $\gamma^k$\\
\STATEnonum \textbf{Initialize} $\{\theta^k_0\}_{k \in \mathcal{U}}$ and $\{Q_n^k(0)\}_{k\in \mathcal{U}, n \in \mathcal{N}}$.\\
\FOR{$f=0 \dots $}
    \STATE set $\{N_f^k=0\}_{k=1}^{K}$ \text{and} $\{\overline{L}_f^k=0\}_{k=1}^{K}$\\
    \FOR{$t=fS+1,fS+2,\dots,(f+1)S$}
        \STATE solve problem \eqref{eq:upper_bound}, obtaining $\{I_s^{k*}(t),R_{n,m}^{k*}(t),P_{n,m}^{*}(t)\}_{s \in \mathcal{S}, (n,m) \in \mathcal{E}, k \in \mathcal{U}}$\\
        \FOR {$s \in \mathcal{S}$}
             \FOR {$k \in \mathcal{U}$}
                \IF {$I_{s}^{k*}(t)=1$}
                    \STATE get the DU $\tau^k(T_s^k(t))$ at the head of queue $Q_{s}^{k}(t)$%\tikzmark{top} \tikzmark{right}\\
                    \STATE produce a decision $\mathcal{C}_s(\tau^k(T_s^k(t)),\theta(f))$\\ 
                    \STATE evaluate loss $L_t^k=L_s(\tau^k(T_s^k(t)),\theta^k_f)$ \\
                \STATE update the average loss $\overline{L}_f^k=\frac{N_f^k}{N_f^k+1}\overline{L}_f^k+\frac{L_t^k}{N_f^k+1}$
                    \STATE update the number of decisions $N_f^k=N_f^k+1$
                \ENDIF
             \ENDFOR
        \ENDFOR
        \STATE update all the system queues $\{\{Q_{n}^{k}(t+1)\}_{n=1}^N\}_{k=1}^K$ via \eqref{eq:system_queues}
    \ENDFOR
    \STATE update the hyperparameters $\{\theta^k_{f+1}\}_{k=1}^{K}$ using \eqref{eq:hyperparameters_update}, \hspace{6pt}
\ENDFOR
%\AddNote{top}{bottom}{right}{A-CRC}
%\AddNote{top1}{bottom1}{right}{LO}
\end{algorithmic}
\caption{Conformal Lyapunov Optimization (CLO)}
\label{alg:clo}
\end{algorithm}

%\vspace{-10pt}
\section{Theoretical Guarantees}\label{sec:theoretical_guarantees}

\begin{figure*}[ht]
     \centering
     \begin{subfigure}[b]{0.36\textwidth}
         \centering
         \includegraphics[width=\textwidth]{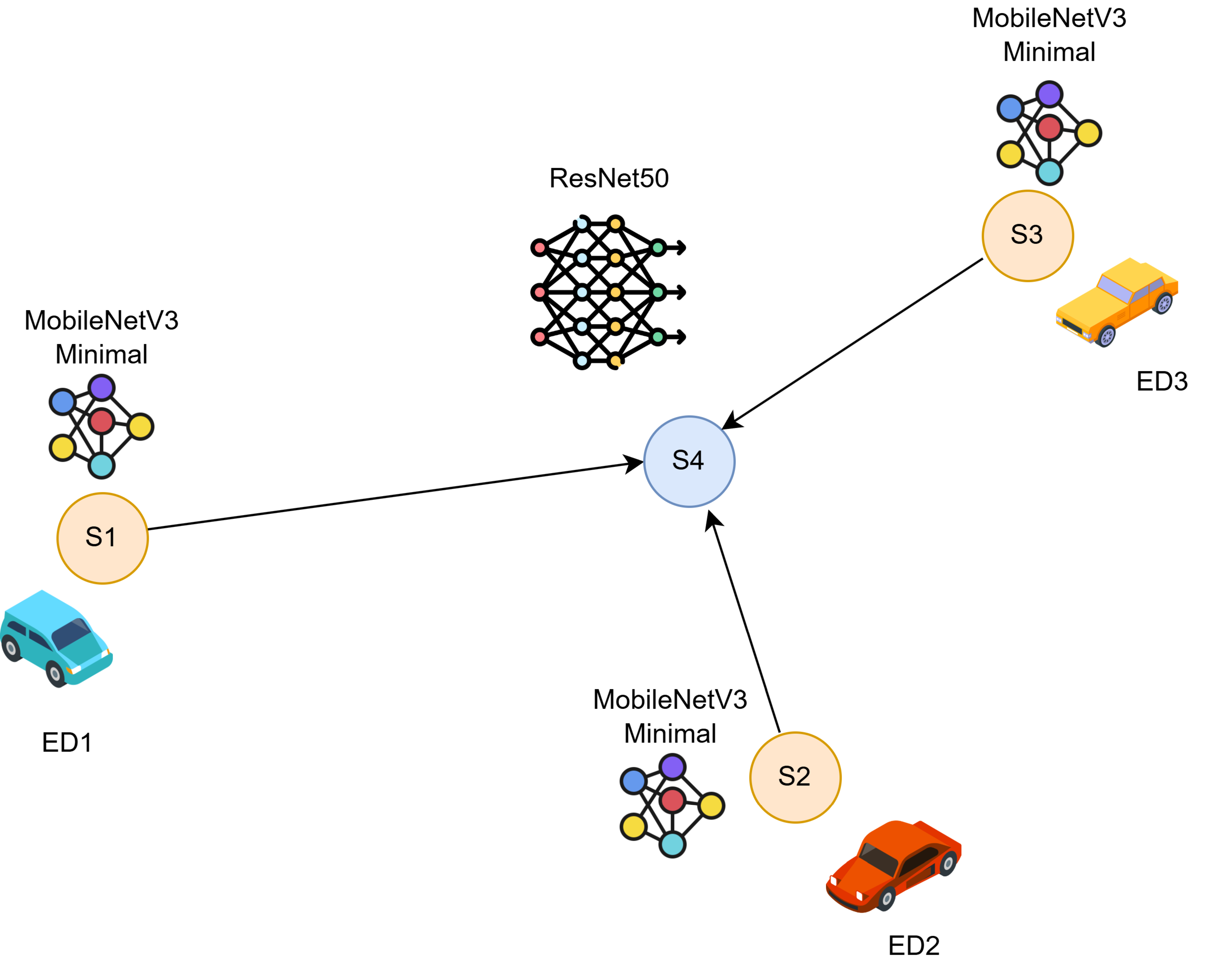}
         \caption{Single-hop network topology.}
         \label{fig:single-hop-network}
     \end{subfigure}
     \hfill
     \begin{subfigure}[b]{0.36\textwidth}
         \centering
         \includegraphics[width=\textwidth]{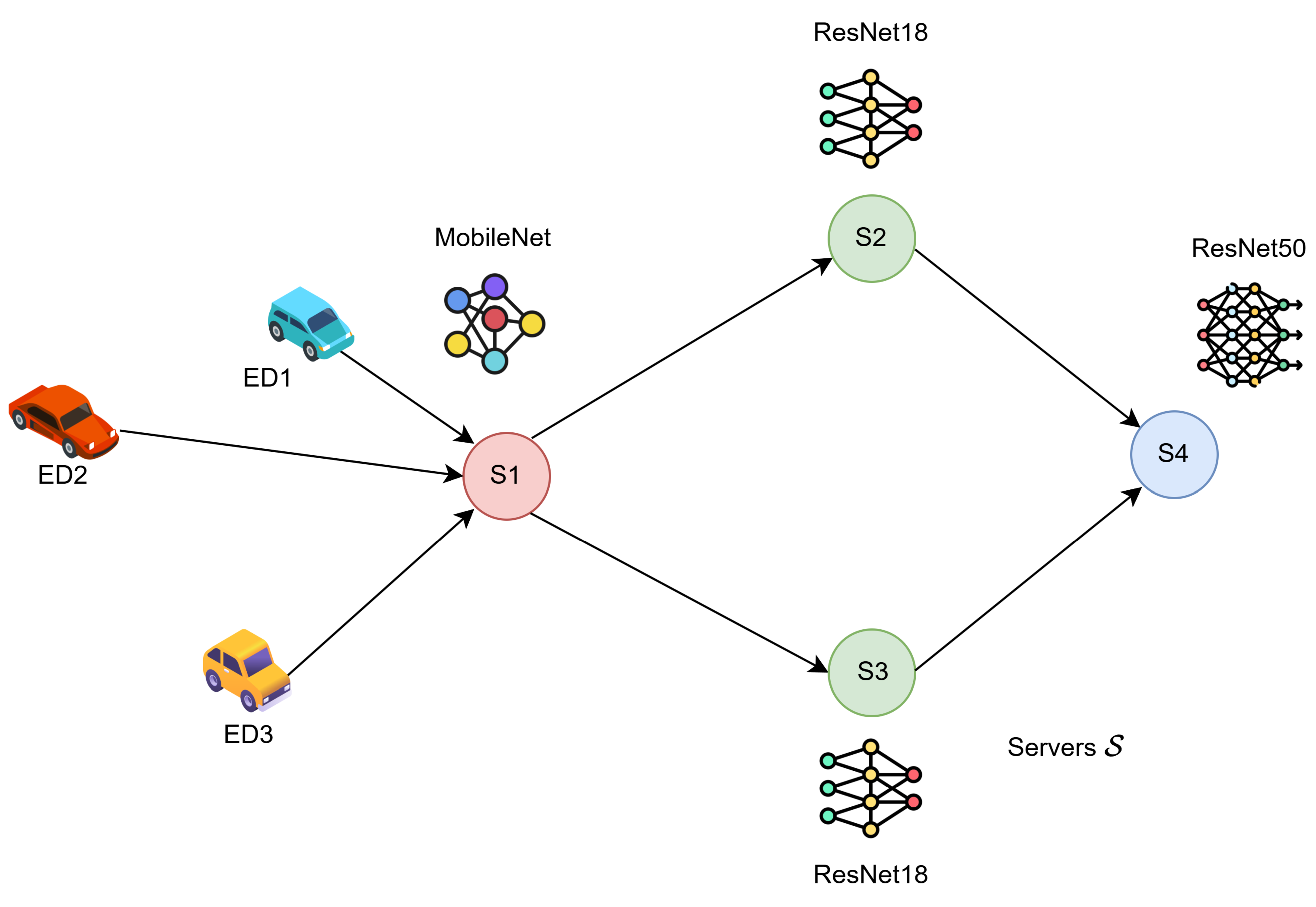}
         \caption{Multi-hop network topology.}
         \label{fig:multi-hop-network}
     \end{subfigure}
     \hfill
        \caption{Network topologies considered in the experimental evaluation.}
    \label{fig:network_topology}
\end{figure*}

In this section, we report  theoretical guarantees for  the proposed CLO protocol. To this end, we first consider the long-term reliability constraint (\ref{eq:long_term_prob}a). The proof of the following claim  is given in Appendix \ref{sec:prof_prop_1}  following reference \cite{feldman2023achieving}.
 % \color{black} As in \cite{feldman2023achieving}, we make the assumption that the thresholds are upper bounded as \begin{equation}\label{ass:theta_lim}\theta^k_f+\gamma^k\leq M \end{equation} for some value $M \in [0,1]$ for all times $t$. In practice, the upper bound $M$ can be set to 1 if no further information is available on the threshold sequence $\theta^k_f$.  To obtain a tighter upper bound in \eqref{eq:clo_upper_bound}, the upper bound $M$ on the thresholds  can be also estimated after the execution of CLO.
\begin{proposition}
\label{prop:rrc_lyap}
Under Assumptions 1 and 2, as the number of frames, $F$, grows large, the deterministic long-term reliability constraint (\ref{eq:long_term_prob}a) is satisfied by CLO for each realization of the stochastic process $\mathbf{\Omega}(t)=\{\mathbf{A}(t),\mathbf{S}(t),\mathbf{T}(t)\}$. {\color{black} Specifically, the following lower and upper bounds $l(m)$, $U(M)$, are satisfied by the average reliability loss \eqref{eq:reliability_const} for any number $F$ of frames 
\begin{align}
    l(m) &= r^k - \textcolor{black}{\frac{r^k d^k}{F}} + \frac{m - \gamma^k - \theta^k_0}{\gamma^k F}, \notag \\
    U(M) &= r^k + \frac{M + \gamma^k - \theta^k_0}{\gamma^k F} + \textcolor{black}{\frac{d^k (1 - r^k)}{F}}, \notag \\
     \quad 
    l(m) &\leq \frac{1}{F}\sum_{f=0}^{F-1}\overline{L_f^k} \leq U(M). \label{eq:clo_upper_bound}
\end{align}
% \begin{equation}\label{eq:clo_upper_bound}
%     \underbrace{r^k-\frac{r^kd^k}{F}+\frac{m-\gamma^k-\theta^k_0}{\gamma^k F}}_{l(m)} \leq \frac{1}{F}\sum_{f=0}^{F-1}\overline{L_f^k}\leq \underbrace{r^k+\frac{M+\gamma^k -\theta^k_0}{\gamma^kF}\color{black}{+{\frac{d^k(1-r^k)}{F}}}}_{U(M)},  
% \end{equation}  
where $M=\max_{f}\{\theta^k_f\}-\gamma^k$ and $m=\min_{f}\{\theta^k_f\}+\gamma^k$.}
\end{proposition}

For classification and binary segmentation tasks, we can set $M=1$ and $m=0$ if no further information is available \cite{feldman2023achieving}. Otherwise, the value of $M$ ($m$) can be estimated from the maximum (minimum) hyperparameter $\theta^k_f$ observed \emph{after} the execution of CLO, thus obtaining tighter (a posteriori) bounds. 
% , obtaining an a posteriori estimation $M_{\mathrm{post}}$. These bounds, which are deterministically respected by CLO, can be used to define an outage probability to assess the performance loss of competitive resource allocation strategies with respect to CLO
% \begin{equation}\label{eq:outage_probability}
% P_{\mathrm{out}}^k(F,M)=\mathbb{P}\left(\frac{1}{F}\sum_{f=1}^{F}\overline{L_{f}^k}>U(M,F)\right),
% \end{equation}
% which quantifies the probability that the bound defined by the constant $M$ is violated at the $F$-th frame. 

% To ease the notation we define $P^k_{\mathrm{out},\mathrm{worst}}(F)=P^k_{\mathrm{out}}(F,1)$ and $P^k_{\mathrm{out},\mathrm{post}}(F)=P^k_{\mathrm{out}}(F,M_{\mathrm{post}})$. 

Proposition \ref{prop:rrc_lyap} shows that, in terms of the reliability constraint, it is advantageous to choose a number of slots per frame, $S$, as small as possible, so as to increase the number of frames $F$ for any given total number of slots  $T=FS$. However, it will be observed next that larger values of $S$ are beneficial to reduce the average cost. 
% As mentioned in Sec.\ \ref{sec:problem_definition}, for a given time horizon $T$, the number of frames is given by $F = T/S$. This proposition clearly indicates that, to minimize the time required to satisfy the reliability constraint, the frame size $S$ should be chosen such that $S \to 1$.

The analysis of the cost function in (\ref{eq:long_term_prob}), and of the average stability constraint (\ref{eq:long_term_prob}a), requires the following standard statistical assumption.

\begin{assumption}\label{ass:iid_ass}
    The process $\mathbf{\Omega}(t)=\{\mathbf{A}(t),\mathbf{S}(t),\mathbf{T}(t)\}$ is i.i.d. over time slots.
\end{assumption}

\begin{proposition}
\label{prop:lyapunov_opt}
Let 
\begin{equation}
    G(t)=\sum_{n=1}^{N}\sum_{k=1}^{K}Q_n^k(t)^2
\end{equation}
be the Lyapunov function for the system's queues, and assume the condition $\mathbb{E}\{G(fS+1)\}\leq\infty$. Under Assumption \ref{ass:iid_ass}, denoting by $J_f^{*}$ the minimum time-average cost at the $f$-th frame achievable by any policy that meets constraint (\ref{eq:long_term_prob}b), CLO satisfies the following properties:
\begin{equation}
    \begin{split}
        &\text{(i)}\; \frac{1}{T}\sum_{t=1}^{T}\mathbb{E}\{J(t)\}\leq \frac{1}{F}\sum_{f=0}^{F-1}\left[J_{f}^{*}+O\left(\frac{1}{S}\right)\right]+\frac{\mu}{V}\\
        &\text{(ii) constraint (\ref{eq:long_term_prob}b) is satisfied,}
    \end{split}
\end{equation}
where $\mu$ is a constant term, and $V$ is the LO hyperparameter that trades the minimization of the objective function (i.e., energy and precision loss) for the average system delay.
\end{proposition}

The role of the Lyapunov function, as well as the proof of this results, are detailed in Section I of the supplemental materials and Appendix \ref{sec:lyapunov_opt_proof}, respectively. This proposition shows that CLO can attain a close-to-optimal performance in the long-run, while satisfying all the constraints in problem (\ref{eq:long_term_prob}). In particular, the sub-optimality of the solution is bounded by a term of the order $\mathcal{O}(1/S)$. Therefore,  improving the network cost requires increasing the frame size $S$.

Overall, the results in this section outline a trade-off in the choice of the number $S$ of slots per frame. In fact, a larger value of $S$ helps obtaining lower levels for the cost function (\ref{eq:long_term_prob}), while smaller values enhance the speed at which the reliability target \eqref{eq:reliability_const} is attained.

\section{Simulation Results}\label{sec:Simulation Results}

\begin{figure*}[ht]
     \centering
          \begin{subfigure}[t]{0.40\textwidth}
         \centering
         \includegraphics[height=0.8\textwidth]{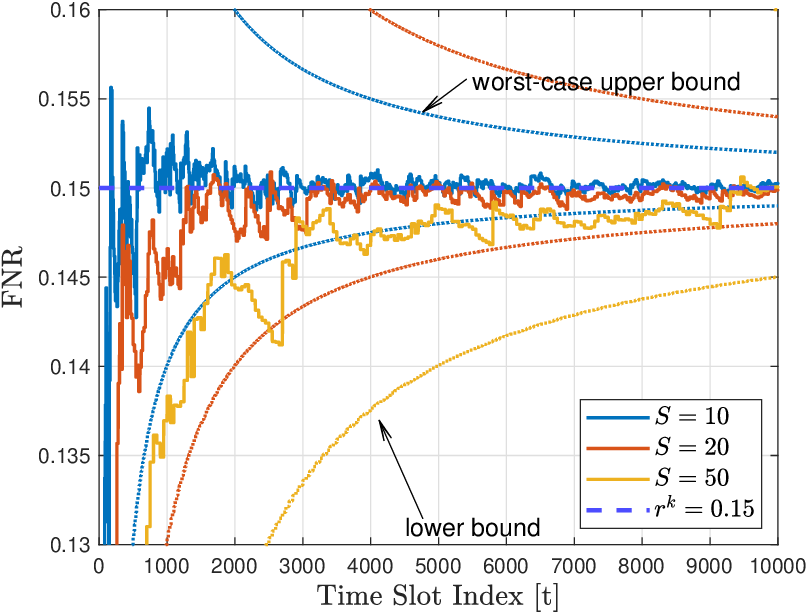}
         \caption{FNR \eqref{eq:reliability_const} as a function of the time slot index for different frame sizes $S$ $(r^k=0.15, \textcolor{black}{d^k=0}, \eta=0.5)$.}
         \label{fig:time_to_reliability}
     \end{subfigure}
     \hfill
     \begin{subfigure}[t]{0.40\textwidth}
         \centering
         \includegraphics[height=0.8\textwidth]{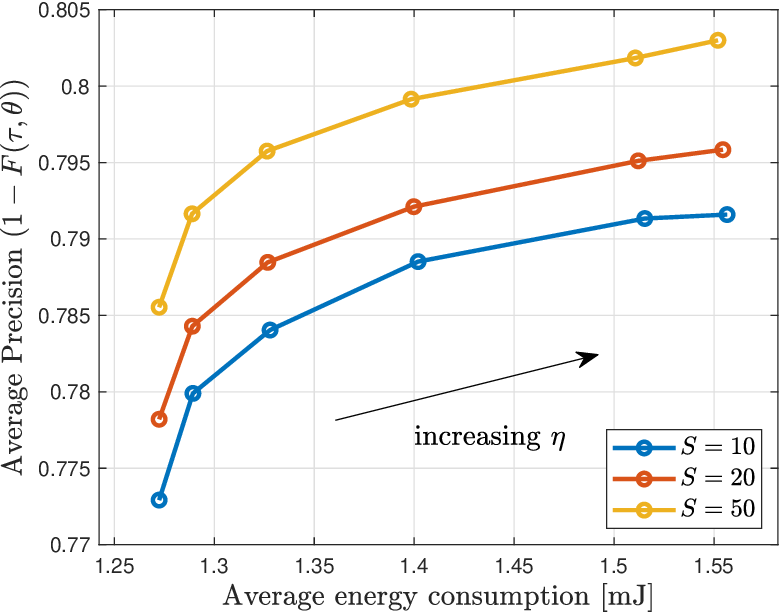}
         \caption{Energy vs. precision trade-off for different frame sizes $S$ \textcolor{black}{($d^k=0$)}.}
         \label{fig:energy_imprecision_trade_off}
     \end{subfigure}
     \hfill
        \caption{FNR evolution and average energy vs. precision trade-off for CLO.}
        \label{fig:first_set_of_segmentation_results}
\end{figure*}

\begin{figure}[ht]
    \centering
    \includegraphics[width=0.85\linewidth]{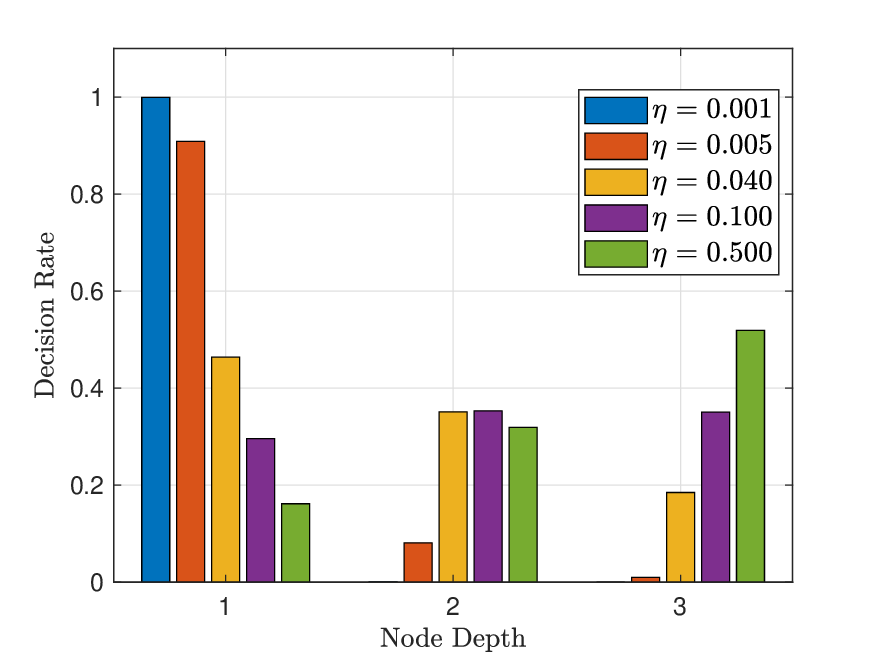}
    \caption{Percentage of decisions at different network nodes for different precision-energy trade-off parameter $\eta$ in \textcolor{black}{\eqref{eq:cost_function}, ($S=50$, $d^k=0$)}.}
    \label{fig:decision_nodes_vs._eta}
\end{figure}

\begin{figure*}[t]
    \centering
     \begin{subfigure}[ht]{0.40\textwidth}
         \centering
         \includegraphics[width=\textwidth]{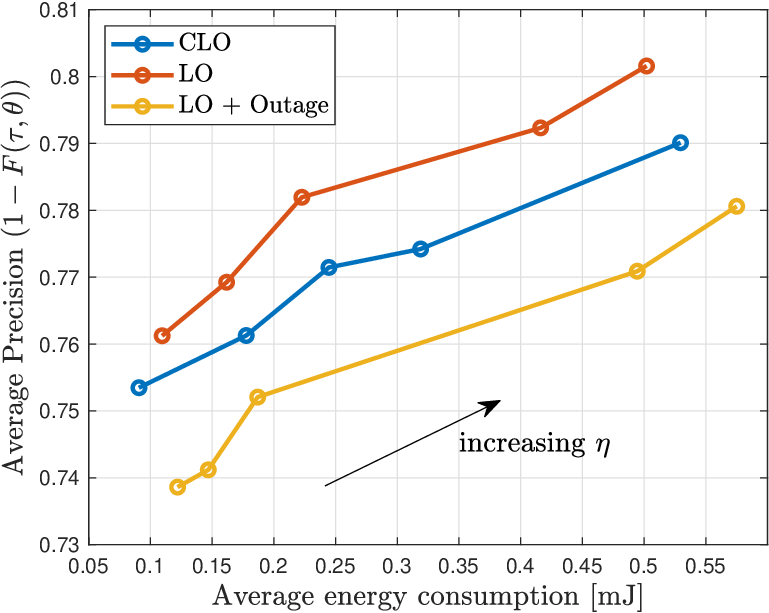}
         \caption{\textcolor{black}{Energy vs. precision trade-off for CLO, LO, and LO with outage probability constraints.}}
         \label{fig:energy_precision_trade_off_lo}
     \end{subfigure}
     \hfill
     \begin{subfigure}[ht]{0.40\textwidth}
         \centering
         \includegraphics[width=\textwidth]{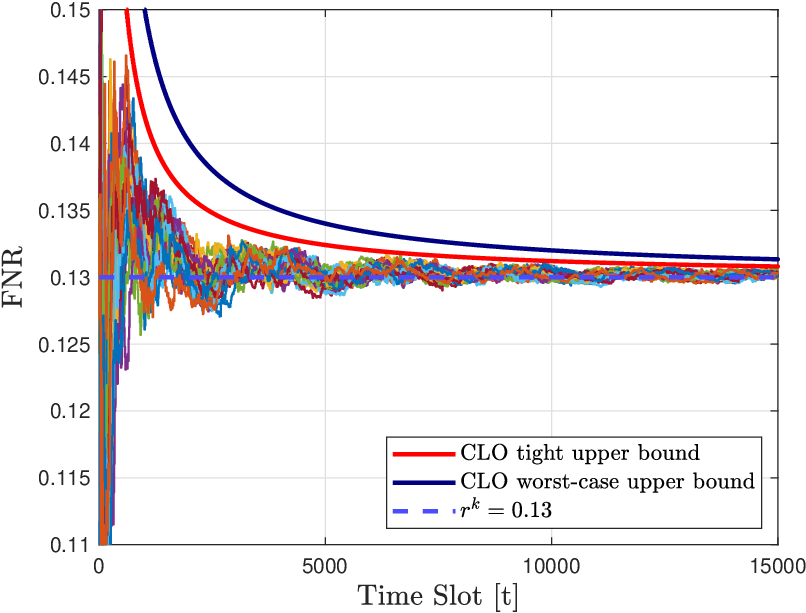}
         \caption{FNR \eqref{eq:reliability_const} as a function of the time slot index for CLO ($\eta=1$, $r^k=0.13$).}
         \label{fig:long_term_conformal}
     \end{subfigure}
     \hfill
          \begin{subfigure}[ht]{0.40\textwidth}
         \centering
         \includegraphics[width=\textwidth]{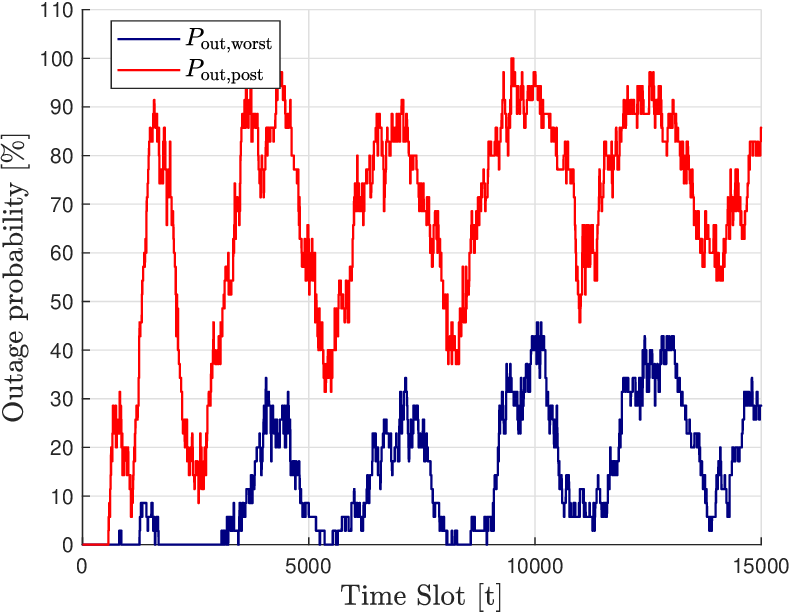}
         \caption{LO outage probabilities  for the worst-case $(M=1)$ and the estimated ($M_{\mathrm{post}}\approx0.6$) upper bounds in \eqref{eq:clo_upper_bound}.}
         \label{fig:long_term_lyapunov_outage}
     \end{subfigure}
     \hfill
     \begin{subfigure}[ht]{0.40\textwidth}
         \centering
         \includegraphics[width=\textwidth]{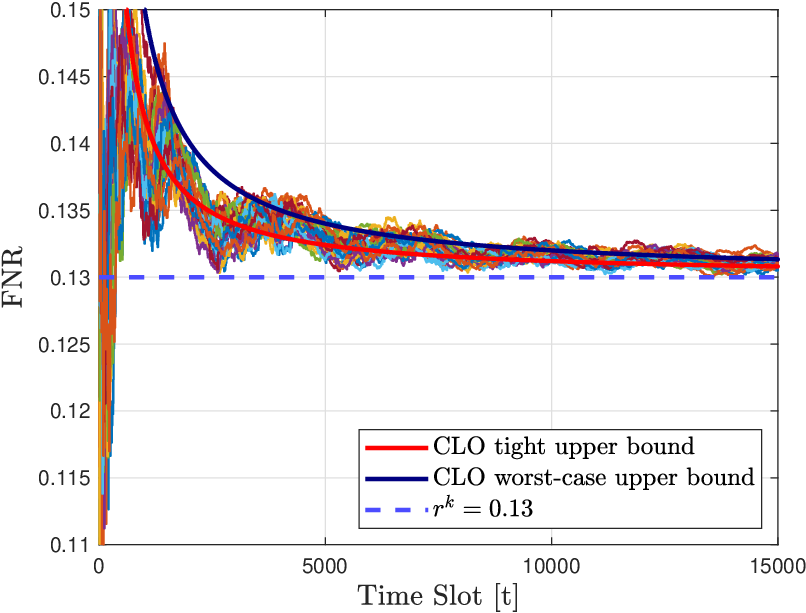}
         \caption{FNR \eqref{eq:reliability_const} as a function of the time slot index for LO ($\eta=1$, $r^k=0.13$).}
         \label{fig:long_term_LO}
     \end{subfigure}
    \caption{Comparisons between LO and CLO \textcolor{black}{assuming no delay in estimation and dissemination of frame-loss information (i.e., $d^k=0$)}: (a)  Energy vs. precision trade-offs for LO and CLO; (b)  Long-term FNR over time for CLO; (c) Average outage probability over time for CLO; (d) Average FNR over time for LO. }
    \label{fig:constraint_satisfaction_comparison}
\end{figure*}

In this section we provide numerical results to test the effectiveness of the proposed CLO protocol and to validate the theoretical guarantees claimed in Section \ref{sec:theoretical_guarantees}.

\subsection{Setting}\label{sec:net_setting}
We consider both a single-hop and a multi-hop network, as summarized in Figure \ref{fig:network_topology}. The single-hop network in Figure \ref{fig:single-hop-network} comprises $K=3$ EDs connected to a single centralized ES, which is equipped with a ResNet50 encoder. Each ED acts also as an ES running a UNet segmentation network \cite{ronneberger2015u} based on a minimal (e.g., low complexity) MobileNetV3 (MNV3) encoder \cite{Iakubovskii:2019}.

In contrast, in the multi-hop architecture shown in Figure \ref{fig:multi-hop-network}, there are  $\vert\mathcal{U}\vert=3$ EDs and $\vert\mathcal{S}\vert=4$ servers. The edge and cloud servers are equipped with UNet inference models, characterized by an increasing complexity (and possibly higher precision), as we move from the EDs towards the edge/cloud servers, employing MobileNetV3, ResNet18 and ResNet50 encoders. 
The NNs employed at each node for image segmentation, along with their computational complexities, are reported in Tables \ref{tab:learning_models_single_hop} and \ref{tab:learning_models_multi_hop}. Their implementation exploits the PyTorch Image Models repository \cite{Wightman_PyTorch_Image_Models}.

\begin{table}[ht]
    \centering
    \begin{tabular}{|c|c|c|}
         \hline
         Nodes & Model Type & Complexity [GMACs]\\
         \hline
         S1,S2,S3 & MNV3 Minimal& 2.55\\
         \hline
         S4 & ResNet50 & 10.63\\
         \hline
    \end{tabular}
    \caption{Segmentation models for the single-hop network.}
    \label{tab:learning_models_single_hop}
\end{table}

\begin{table}[ht]
    \centering
    \begin{tabular}{|c|c|c|}
         \hline
         Nodes & Model Type & Complexity [GMACs]\\
         \hline
         S1 & MNV3 Large & 3.06\\
         \hline
         S2, S3 & ResNet18 & 5.39\\
         \hline
         S4 & ResNet50 & 10.63\\
         \hline
    \end{tabular}
    \caption{Segmentation models for the multi-hop network.}
    \label{tab:learning_models_multi_hop}
\end{table}

The links between nodes are assumed to be wireless and, for simplicity, characterized by a Rayleigh distribution with the same average path-loss $\mathrm{PL}=90 \hspace{2pt} \mathrm{dB}$. We set a maximum transmit power $P_n^{\mathrm{max}}=3.5 \hspace{2pt}\mathrm{W}$ for all the nodes $n \in \mathcal{N}$, and the same noise power spectral density $N_0=-174 \hspace{2pt} \mathrm{dBm/Hz}$. All the links are characterized by the same transmission bandwidth $B_{n,m}=20$ MHz  for all $(n,m) \in \mathcal{E}$. We set a time time slot duration $\delta=50$ ms, corresponding to the channel coherence time. \textcolor{black}{We assume that there is no delay associated with the estimation and dissemination of frame loss information, i.e., \( d^k = 0 \) for all EDs (see Appendix D for further results).}
The per-slot problem \eqref{eq:upper_bound} is solved using the Python-based CVX implementation (CVXPY) \cite{diamond2016cvxpy,agrawal2018rewriting}.

\subsection{Task Description}

We focus  on a binary image segmentation task, with images and binary object masks obtained from the Cityscapes dataset of urban scenarios \cite{Cordts_2016_CVPR}. We split this data set in 10,000 images for training, and 10,000 images for testing the segmentation NNs. The images are resized to $256\times256\times3$ pixels and encoded in a 32-bit format, resulting in an image size of $W^k=768$ KB. Since the dataset is originally designed for multi-class semantic segmentation, we formulate the task as a binary segmentation by labeling only car-related pixels as segmentation objective, treating all the others as background. For instance, this task could be useful in vehicular applications.

We set $r^k = 0.15$ in \eqref{eq:reliability_const} for the FNR constraint, while for the precision loss $F_s(x,\theta)$ we consider the ratio of the pixels falsely identified as part of the car, over the true ones \footnote{\textcolor{black}{We chose this precision measure because, unlike \eqref{eq:precision}, it is relative to the size of the object of interest, making it more meaningful for objects that are significantly smaller than the background.}}, i.e., 
\begin{equation}\label{eq:precision_metric}
F_s(x,\theta)=\min\left(\frac{\abs{\overline{y}_{\mathrm{true}}}\cap\mathcal{C}(x,\theta)}{|y_{\mathrm{true}}|},1\right).
\end{equation}

To predict the precision loss $F_s(x,\theta)$ (see Section \ref{subsec:model}), we consider a set of low-complexity NNs based on the PSPNet architecture \cite{zhao2017pyramid}. Each NN approximates the precision loss $F_s(x,\theta)$  of the inference model employed at the $s$-th ES, which are summarized in Tables I and II. These NNs are trained using knowledge distillation \cite{gou2021knowledge}, by minimizing a linear combination of the segmentation loss (i.e., cross-entropy) and the Kullback–Leibler divergence between the outputs of the teacher and student NNs. This approach enables the student NNs to replicate the segmentation masks produced by the teacher NNs, i.e., the models actually deployed at the EDs and ESs. From Tables \ref{tab:learning_models_single_hop}, \ref{tab:learning_models_multi_hop}, and \ref{tab:precision_loss_approximators}, we observe that the complexity of the precision predictors (PP) takes values in the range $2\%-7\%$ of the complexity of the actual segmentation models. \footnote{Thus the use of the PPs at the (unique) control center makes sense because it requests a much lower complexity than directly performing the segmentation assigned to the ESs.} 

% Note that even though each PP provides an estimate of the precision accurate enough to drive the optimization strategy, it does not necessarily produce a very accurate segmentation mask.}

\begin{table}[ht]
    \centering
    \begin{tabular}{|c|c|c|}
         \hline
         Model Type & Approximator & Complexity [MMACs]\\
         \hline
         MNV3 Minimal & MNV3 Minimal & 50.90\\
         \hline
         MNV3 Large & MNV3 Large & 140.25\\
         \hline
         ResNet18, ResNet50 & MobileOne S0 & 783.50\\
         \hline
    \end{tabular}
    \caption{Complexity in terms of milions of multiplications and accumulations (MMACs) operations for the approximation models used to estimate the imprecision function.}
\label{tab:precision_loss_approximators}
\end{table}

\subsection{Precision-Reliability Trade-Off}\label{sec:prec_rel_to}
We start by validating the theoretical guarantees presented in Proposition \ref{prop:rrc_lyap} and Proposition \ref{prop:lyapunov_opt}, by assessing the impact of a different frame size $S$ on the trade-offs between energy consumption, precision, and reliability. 

We consider the multi-hop network depicted in Figure \ref{fig:multi-hop-network}. For all the users the learning rates are set to $\gamma^k=0.5$, and the initial segmentation thresholds are set as $\theta^k_0=0.5$. Without restriction of generality, we trade the function cost for average delay employing a Lyapunov trade-off parameter $V=\num{2e2}$ (cf. \eqref{eq:upper_bound}), and a set of energy-precision trade-off parameters $\eta \in \{0.1,0.5,1,2,4,5\} \times 10^{-1}$ (cf. \eqref{eq:cost_function}). The environment is assumed to be stationary, with inference tasks $\tau^k(t)$ that are i.i.d., and generated according to a Bernoulli distribution, with a probability $\lambda_k= 0.5$ for all the users.

Figure \ref{fig:time_to_reliability} plots the FNR evolution in time for different frame sizes $S$, where each curve is obtained for the same single realization of tasks. \textcolor{black}{The figure shows the theoretical deterministic guarantees} offered by CLO.

% which are satisfied for every realization. 

Figure \ref{fig:energy_imprecision_trade_off} shows the trade-off between overall average precision, evaluated as $1 - F_s(x,\theta)$, and the transmission energy consumption of all the nodes. \textcolor{black}{The curves are obtained by varying the penalty $\eta$ } in \eqref{eq:cost_function} and  by averaging over the last 1,000 time slots (of the total $T=10,000$), as well as over \textcolor{black}{$30$} different realizations of tasks. Increasing the average precision requires offloading computations to network nodes farther away from the users, increasing the transmission energy consumption. This is confirmed by Figure \ref{fig:decision_nodes_vs._eta}, which shows how the nodes decision percentages vary with the nodes depth, for different values of the precision-energy trade-off parameter $\eta$. 

Figures \ref{fig:time_to_reliability} and \ref{fig:energy_imprecision_trade_off} confirm that, in accordance to Propositions \ref{prop:rrc_lyap} and \ref{prop:lyapunov_opt}, increasing the frame size $S$ allows CLO to attain a higher precision over a finite time duration, but a slower convergence to the FNR target value.

\textcolor{black}{\subsection{Comparison with LO-based Resource Allocation Strategies}
\begin{figure}
    \centering
    \includegraphics[width=0.90\linewidth]{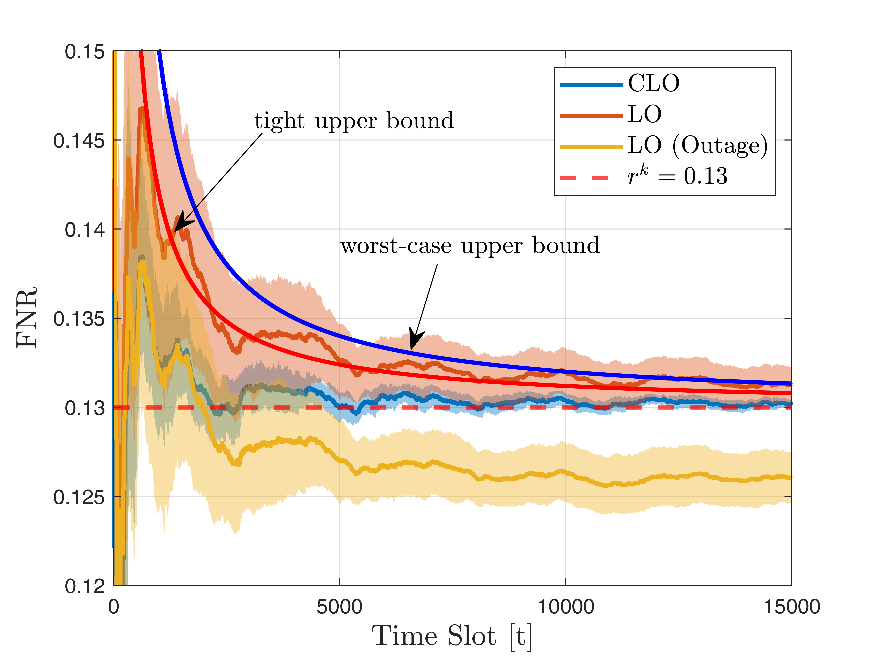}
    \caption{\textcolor{black}{Long-term FNR reliability loss for CLO, LO, and LO with outage probability constrains} \color{black}{$(\eta=1, r^k=0.13, d^k=0)$.}}
    \label{fig:long-term-reliability-comparison}
\end{figure}}

We compare the performance of the proposed CLO scheme with \textcolor{black}{resource allocation strategies based on the conventional LO framework. Recall that LO addresses only long-term constraints characterized by averages or higher-order statistical moments. Two primary baseline strategies are considered: 1) standard LO strategies tailored to long-term average constraints~\cite{neely2022stochastic}, and 2) LO strategies designed to handle outage probability constraints~\cite{merluzzi2020dynamic}.
In the standard LO, we replace the \emph{deterministic} constraint \eqref{eq:reliability_const} by the average constraint \cite{neely2022stochastic}  
\begin{equation}\label{eq:average_reliability}
    \lim_{F\to\infty}\frac{1}{F}\sum_{f=0}^{F-1}\mathbb{E}\left\{\overline{L_f^k}\right\}\leq r^k.
\end{equation} 
For the second benchmark, we impose the following outage probability constraint 
\begin{equation}\label{eq:long_term_outage_constraint}
    \lim_{F\to\infty}\frac{1}{F}\sum_{f=0}^{F-1}\mathbb{P}_r(\overline{L^k_f} > L^k_{\max}) \leq \epsilon^k,
\end{equation}
where $L^k_{\max}$ denotes the maximum tolerable reliability per frame, and $\epsilon^k$ specifies the target long-term outage probability.
}
\textcolor{black}{LO guarantees long-term reliability constraints by reformulating them as queue stability conditions associated with virtual queues for \eqref{eq:average_reliability}, and \eqref{eq:long_term_outage_constraint} \cite{neely2022stochastic}. We refer to Section II of the supplementary materials for further details.
}

Accordingly, while CLO updates the reliability hyperparameters $\theta^k$ at the end of each time frame (every $S$ time slots), the competitive LO formulations treat them as variables to be optimized at each time slot $t$. Treating these variables as discrete within the set $\{0.1, 0.2, \dots, 0.9\}$ yields a mixed-integer optimization problem, whose complexity grows exponentially with the number $K$ of users. Thus, to make the LO problems computationally feasible, in each slot we force all the users to employ the same threshold, i.e., $\theta^k(t) = \theta^*(t)$. 

We consider the single-hop network architecture shown in Figure \ref{fig:single-hop-network}, with an i.i.d. generation of new tasks according to a Bernoulli distribution with probability $\lambda^k$ for any user. We simulate a non-stationary environment, where $\lambda^k\in \{0.4,0.8\}$ may switch every $100$ slots, with a probability $p=0.5$. The Lyapunov and penalty trade-off parameters are $V=\num{2e2}$, and $\eta \in \{\num{1e-2},\num{5e-2},\num{1e-1},\num{5e-1},1\}$. The frame size for CLO is $S=10$. To make fair comparisons between LO and CLO, we set for LO a virtual queue step size $\beta^k=0.5$, which is equivalent to the CLO learning rate $\gamma^k=0.5$.
\textcolor{black}{\underline{\textit{Comparisons with LO with Average Reliability Control: }}}The main reason to compare LO and CLO is understanding the price inevitably incurred by CLO to guarantee a \emph{deterministic}, per-realization, reliability constraint. To this end,  Figure \ref{fig:energy_precision_trade_off_lo} compares the average  precision  achieved by LO \textcolor{black}{with average reliability constraints} and by CLO versus the energy consumption. The results are evaluated at convergence of the reliability constraint, by averaging over the last $1000$ of $T=15,000$ time slots. LO \textcolor{black}{with average reliability constraints} is observed to achieve a higher precision for the same energy consumption as compared to CLO, with the gap quantifying the cost paid by CLO to ensure \emph{deterministic} reliability constraints.

The reliability constraints are highlighted by a dotted blue line in Figures  \ref{fig:long_term_conformal} and \ref{fig:long_term_LO}, which plot the FNR evolution versus time for \textcolor{black}{30} tasks realizations. These plots are obtained under a comparable energy consumption for the two optimization strategies, which corresponds to the rightmost points in Figure \ref{fig:energy_precision_trade_off_lo}. The continuous blue curves in Figures \ref{fig:long_term_conformal} and \ref{fig:long_term_LO} highlight the worst-case FNR upper bound, computed by setting $M=1$ in \eqref{eq:clo_upper_bound}, while the red curves identify an (a posteriori) upper bound, obtained by estimating the value of the constant $M$ among \textcolor{black}{30} realizations of CLO.

Figure \ref{fig:long_term_lyapunov_outage} shows two FNR outage curves for LO, defined as the probabilities to violate during convergence the worst case and the a posteriori upper bound, of CLO. The curves are obtained by evaluating the fraction of realizations (among 50), with an FNR value above the upper bound in Proposition 1, \emph{deterministically} guaranteed by CLO for any realization. While CLO consistently remains within the theoretical bounds, LO exhibits a high likelihood to exceed them, with a probability that increases over time as the bound gets tighter.
\textcolor{black}{\underline{\textit{Comparisons with LO with Outage Probability Control: }} To ensure a fair comparison with LO strategies that incorporate outage probability control, we set the threshold value as \( L^k_{\max} = r^k + r^k/10\) in \eqref{eq:long_term_outage_constraint}, corresponding to a 10\% margin above the target reliability \( r^k = 0.13 \), for all users. With this choice, we evaluated the empirical outage probability achieved by CLO, defined as the frequency with which \( \overline{L^k_f} \) exceeds \( L^k_{\max} \), averaged over 30 independent realizations. The resulting average was \( \epsilon^k \approx 32\% \), which was then adopted as the outage probability target in the outage constraint~\eqref{eq:long_term_outage_constraint}.}

\textcolor{black}{Figure~\ref{fig:energy_precision_trade_off_lo} illustrates that the LO strategy with outage probability constraints (yellow curve) yields the worst energy--precision trade-off among the compared methods. This result stems from the fact that, in order to satisfy the outage probability constraint, LO tends to prioritize reliability over precision.} 

\textcolor{black}{This observation is corroborated by Figure~\ref{fig:long-term-reliability-comparison}, which presents the long-term reliability achieved by the three competing strategies. The plot shows the average long-term reliability loss, with results averaged over 30 independent realizations of the task sequence. Unlike the standard LO and the proposed CLO scheme, the LO strategy with outage probability control consistently exhibits lower long-term reliability loss. This, due to the intrinsic trade-off between precision and reliability, also leads to diminished precision performance. Furthermore, it is observed that the CLO scheme achieves the lowest standard deviation among all the three strategies, highlighting its advantage in attaining a more stable solution in terms of long-term reliability.
}

\textcolor{black}{
\subsection{Effect of the Trade-off Parameter $V$}
}
\begin{figure*}[ht]
    \centering
     \begin{subfigure}[ht]{0.40\textwidth}
         \centering
         \includegraphics[width=\textwidth]{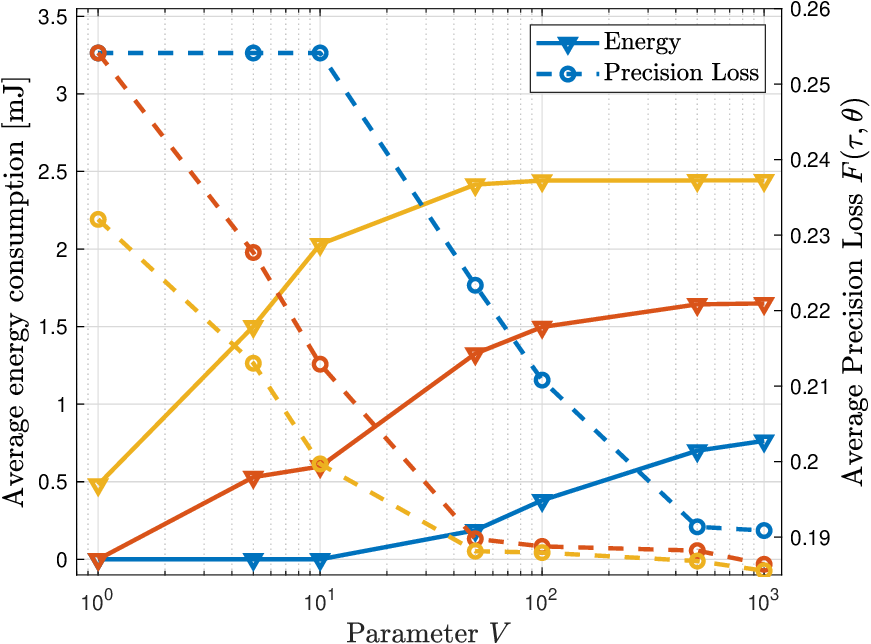}
         \caption{\textcolor{black}{Average energy consumption and precision loss vs. parameter $V$ ($d^k=0$). See Figure \ref{fig:energy_latency_trade_off} for the color legend.}}
         \label{fig:energy_precision_v}
     \end{subfigure}
     \hfill
     \begin{subfigure}[ht]{0.40\textwidth}
         \centering
         \includegraphics[width=\textwidth]{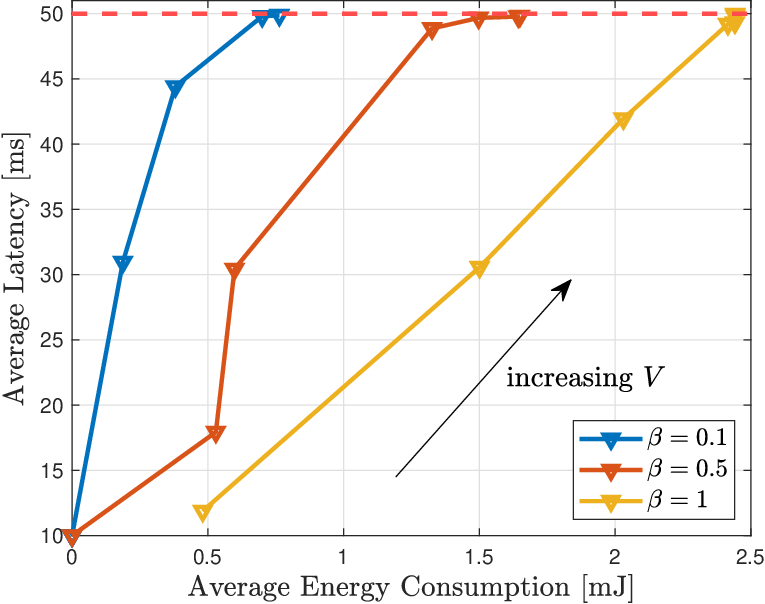}
         \caption{\textcolor{black}{Average energy vs. latency trade-off.\vspace{11pt}}}
         \label{fig:energy_latency_trade_off}
     \end{subfigure}
    \caption{\textcolor{black}{Behavior of CLO for different values of the Lyapunov trade-off parameter $V$: (a) Average energy consumption and precision loss as a function of $V$; (b) Energy vs latency trade-off.}}
    \label{fig:v_analysis}
\end{figure*}
\textcolor{black}{In this section, we investigate the impact of the Lyapunov trade-off parameter $V$ (see~\eqref{eq:upper_bound}) on the network cost under both average latency and strict reliability constraints. As established in Lyapunov optimization theory, the parameter $V$ plays a crucial role in balancing performance and queue stability. Specifically, as the parameter $V$ increases, the average cost achieved by LO deviates from the optimum by an additive error of order $\mathcal{O}(1/V)$, while the average queue size grows proportionally to $\mathcal{O}(V)$ \cite{neely2022stochastic}.
}

\textcolor{black}{To highlight the effect of the trade-off parameter $V$ on CLO, we consider the single-hop edge-inference scenario depicted in Figure \ref{fig:single-hop-network} with edge-to-edge latency constraints. Following the methodology proposed in~\cite{binucci2023multi}, we augment problem~\eqref{eq:long_term_prob} by incorporating an average constraint on the total queue length for each user, defined as
\begin{equation}\label{eq:latency_constraint}
    \lim_{T \to \infty} \frac{1}{T} \sum_{t=1}^{T} \mathbb{E}\{Q^k_{\mathrm{tot}}(t)\} \leq Q^k_{\mathrm{avg}} \hspace{20pt} \forall k,
\end{equation}
where $Q^k_{\mathrm{tot}}(t) = Q^k_k(t) + Q_s^k(t)$, and $Q_s^k(t)$ denotes the queue at the centralized edge server (i.e., node S4 in Figure~\ref{fig:single-hop-network}). This constraint can be interpreted as an average latency constraint. Indeed, assuming a constant task arrival rate $\overline{A^k} = \lambda^k / \delta$ (in tasks/sec) and a stationary queuing system, exploiting Little's Law, the total latency and the average latency constraint can be written as $D^k_{\mathrm{tot}}(t) = Q^k_{\mathrm{tot}}(t) / \overline{A^k}$, and $D^k_{\mathrm{avg}}=Q^k_{\mathrm{avg}}/\overline{A^k}$ respectively.
}

\textcolor{black}{
We consider a simulation time $T=10,000$ slots of $\delta=10$ ms and a reliability loss constraint $r^k=0.15$. Reliability is computed over frames composed of $S=10$ slots.  The inference tasks are encoded with 8 bit per pixel, resulting in a task size $W^k=192$ KB, and generated according to an i.i.d. Bernoulli distribution with probability $\lambda^k=0.8$ for all the users. We impose a queue length constraint $Q^k_{\mathrm{avg}}=4$ tasks/slot, equivalent to a latency constraint $D^k_{\mathrm{avg}}=50$ ms. 
}

\textcolor{black}{
To investigate the impact of the trade-off parameter \( V \) on the precision/reliability balance, we first rewrite the cost function in \eqref{eq:cost_function} as
\begin{equation}
    (1 - \beta)E_{\mathrm{tot}}(t) + \beta F_{\mathrm{tot}}(t), \quad \beta = \frac{\eta}{1 + \eta},
\end{equation}
where the parameter \( \beta \in [0,1] \) regulates the trade-off between energy consumption and precision loss. We evaluate the strategy for $\beta \in \{0.1, 0.5, 1\}$ and $V \in \{1, 5, 10, 50, 100, \dots, 1000\}$.
}

\textcolor{black}{
 Figure~\ref{fig:energy_precision_v} reports the average energy consumption and precision loss as functions of the trade-off parameter $V$. Each curve corresponds to a fixed value of the weighting parameter $\beta$, with $V$ varying across the specified range. Results are obtained by averaging over the last 1{,}000 time slots, after convergence. As $V$ increases, the average precision loss decreases, with a consequent higher energetic consumption. On the other hand, higher values of $\beta$ result in improved precision due to more frequent task offloading to the edge server, which also leads to higher energy consumption. 
}

\textcolor{black}{
 Figure~\ref{fig:energy_latency_trade_off} illustrates the trade-off between energy consumption and latency. It can be observed that the optimization strategy consistently satisfies the average latency constraint, which is indicated by the red dashed line. Specifically, as the trade-off parameter $V$ increases, both energy consumption and latency increase, until the long-term latency constraint is tightly met. This behavior confirms that larger values of $V$ lead to a higher congestion state in the system.
}

\section{Conclusions}
\label{sec:Conclusions}
This paper introduces conformal Lyapunov optimization (CLO), a novel optimization framework that addresses  optimal resource managements for network-based learning, under strict and deterministic constraints on the learning reliability. CLO integrates the standard optimization framework of Lyapunov optimization (LO), with the novel reliability mechanism of online conformal risk control.  Simulation results have validated the theoretical guarantees of CLO in terms of long-term reliability performance,  highlighting its advantages when compared with resource allocation strategies based on LO.

Future research directions may include the exploration of distributed implementations of CLO, as well as applications for more complex scenarios involving multi-carrier transmissions, interfering users, latency, and transmission outages.

\begin{appendices}
\section{Monotonicity Proofs for Reliability and Precision Losses}\label{sec:mon_proofs}

\begin{proof}[Miscoverage and Set-Size Losses]
Let $\theta_1 < \theta_2$, and define the prediction set as $\mathcal{C}(x,\theta) = \{ y \in \mathcal{Y} : p(y|x) \geq \theta \}$. Since $p(y|x) \geq \theta_2$ implies $p(y|x) \geq \theta_1$, it follows that $\mathcal{C}(x,\theta_2) \subseteq \mathcal{C}(x,\theta_1)$. Consequently, $\mathbbm{1}(y_{\mathrm{true}} \notin \mathcal{C}(x,\theta_1)) \leq \mathbbm{1}(y_{\mathrm{true}} \notin \mathcal{C}(x,\theta_2))$, showing that the miscoverage loss is non-decreasing with respect to parameter $\theta$.

For the set-size precision loss, since $\mathcal{C}(x,\theta_2) \subseteq \mathcal{C}(x,\theta_1)$, we have $|\mathcal{C}(x,\theta_2)| \leq |\mathcal{C}(x,\theta_1)|$. Dividing both sides by $|\mathcal{Y}|$ yields
$\frac{|\mathcal{C}(x,\theta_2)|}{|\mathcal{Y}|} \leq \frac{|\mathcal{C}(x,\theta_1)|}{|\mathcal{Y}|}$, proving that the set-size precision loss is non-increasing with respect to parameter $\theta$.
\end{proof}

\begin{proof}[FNR and FPR losses]
Let $\theta_1 < \theta_2$, and define the prediction set as $\mathcal{C}(x,\theta) = \{(i,j) : p(i,j|x) \geq \theta \}$. Since $p(i,j|x) \geq \theta_2$ implies $p(i,j|x) \geq \theta_1$, it follows that $\mathcal{C}(x,\theta_2) \subseteq \mathcal{C}(x,\theta_1)$. For a fixed $y_{\mathrm{true}}$, we have $(y_{\mathrm{true}} \cap \overline{\mathcal{C}(x,\theta_1)}) \subseteq (y_{\mathrm{true}} \cap \overline{\mathcal{C}(x,\theta_2)})$, and consequently, $\frac{|y_{\mathrm{true}} \cap \overline{\mathcal{C}(x,\theta_1)}|}{|y_{\mathrm{true}|}} \leq \frac{|y_{\mathrm{true}} \cap \overline{\mathcal{C}(x,\theta_2)}|}{|y_{\mathrm{true}}|}$, showing that the FNR loss is non-decreasing with respect to the parameter $\theta$.

For the FPR loss, since $\mathcal{C}(x,\theta_2) \subseteq \mathcal{C}(x,\theta_1)$, we have $(\overline{y}_{\mathrm{true}} \cap \mathcal{C}(x,\theta_2)) \subseteq (\overline{y}_{\mathrm{true}} \cap \mathcal{C}(x,\theta_1))$, and hence $\frac{|\overline{y}_{\mathrm{true}} \cap \mathcal{C}(x,\theta_2)|}{|\overline{y}_{\mathrm{true}}|} \leq \frac{|\overline{y}_{\mathrm{true}} \cap \mathcal{C}(x,\theta_1)|}{|\overline{y}_{\mathrm{true}}|}$, proving that the FPR loss is non-increasing with respect to the parameter $\theta$.
\end{proof}

\section{Proof of Proposition \ref{prop:rrc_lyap}}\label{sec:prof_prop_1}
\begin{proof}
\textcolor{black}{
Assuming the presence of a finite estimation and dissemination delay of frame loss information $d^k$, the long-term reliability loss at the $F$-th frame can be written as
\begin{equation}\label{eq:first_sum}
    \frac{1}{F}\sum_{f=0}^{F-1}\overline{L}_f^k=\frac{1}{F}\left[\sum_{f=0}^{F-d^k-1}\overline{L}_f^k+\sum_{f=F-d^k}^{F-1}\overline{L}_f^k\right].
\end{equation}
The first sum in the right hand side of \eqref{eq:first_sum} can be written as
\begin{equation}
    (F-d^k)\left[\frac{1}{F-d^k}\sum_{f=0}^{F-d^k-1}\overline{L}_f^k\right].
\end{equation}
Furthermore, from \cite{feldman2023achieving} the dynamic update of the reliability hyperparameters \eqref{eq:hyperparameters_update}, leads to the following chain of inequalities 
\begin{equation}
    r^k+\frac{m-\gamma^k-\theta_0^k}{(F-d^k)\gamma^k}\leq\frac{1}{F-d^k}\hspace{-6pt}\sum_{f=0}^{F-d^k-1}\hspace{-6pt}\overline{L}_f^k\leq r^k+\frac{M+\gamma^k-\theta_0^k}{(F-d^k)\gamma^k}.
\end{equation}
Thus, multiplying by $(F-d^k)$, and taking into account that the second term in the right hand side of \eqref{eq:first_sum} is always in $[0,d^k]$ thanks to the boundedness assumption on the reliability loss, we end up with the following bounds for the average reliability loss at the $F$-th frame
\begin{equation}
    l(m)\leq \frac{1}{F}\sum_{f=0}^{F-1}\overline{L^k_f}\leq U(m),
\end{equation}
where the bounds are defined as $l(m)=r^k-\frac{r^kd^k}{F}+\frac{m-\gamma^k-\theta_0^k}{\gamma^kF}$, and     $U(M)=r^k+\frac{M+\gamma^k-\theta_0^k}{\gamma^kF}+\frac{d^k(1-r^k)}{F}$.}
\end{proof}

\section{Proof of Proposition \ref{prop:lyapunov_opt}}\label{sec:lyapunov_opt_proof}
\begin{proof}
According to Theorem 4.8 of \cite{neely2022stochastic}, under i.i.d. assumptions on $\mathbf{\Omega}(t)$ LO ensures the following inequality 
\begin{equation}
\begin{split}
    \frac{1}{T}\sum_{t=1}^{T}\mathbb{E}\{J(t)\}&=\frac{1}{F}\sum_{f=0}^{F-1}\frac{1}{S}\sum_{t=fS+1}^{(f+1)S}\mathbb{E}\{J(t)\}\\
    &\leq \frac{1}{F}\sum_{f=0}^{F-1}\left[J_f^{*}+\frac{\mathbb{E}\{G(fS+1)\}}{VS}\right]+\frac{\mu}{V},
\end{split}
\end{equation}
where $\mu$ is a constant term \cite{neely2022stochastic}. Since we are assuming that $\mathbbm{E}\{G(fS+1)\}\leq\infty$, for a fixed value of the penalty parameter $V$, as $S \to \infty$, we end up with an approximate solution whose value is closer to the optimal value of the per-frame resource allocation problem $J_f^*$. Furthermore, by employing the upper-bound presented in Section I of the supplemental materials, and applying theorem 4.8 in \cite{neely2022stochastic} we also ensure that the LDPP function (cf. Section I of the supplementary items) is bounded for each slot $t \in [fS+1,(f+1)S]$ as follows
\begin{equation}
    \Delta_p(t)\leq B+VJ_f^*,
\end{equation}
where $B$ is a constant term. According to Theorem 4.2 in \cite{neely2022stochastic}, this condition ensures the mean-rate stability of all the queues, as requested by constraint  (\ref{eq:long_term_prob}b). 
\end{proof}

\begin{figure*}[ht]
     \centering
          \begin{subfigure}[t]{0.40\textwidth}
         \centering
         \includegraphics[width=\textwidth]{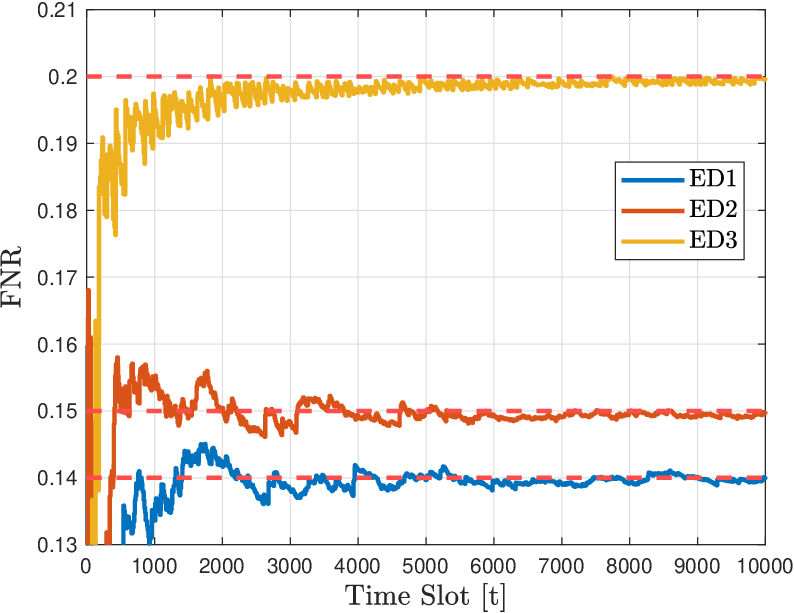}
         \caption{\textcolor{black}{FNR \eqref{eq:reliability_const} as a function of the time slot index for users operating under different reliability constraints $(\eta=0.5)$.}}
         \label{fig:time_to_reliability_users_different_constraints}
     \end{subfigure}
     \hfill
     \begin{subfigure}[t]{0.40\textwidth}
         \centering
         \includegraphics[width=\textwidth]{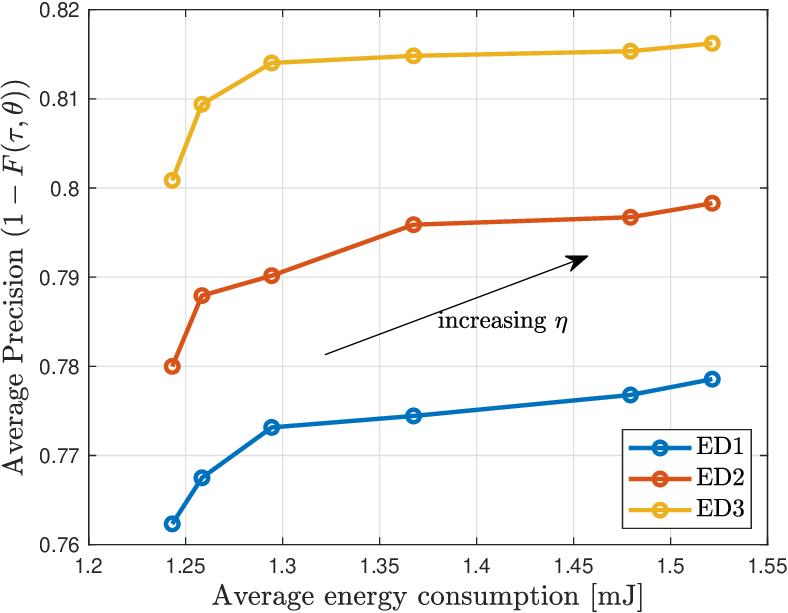}
         \caption{\textcolor{black}{Energy vs. precision trade-off for users operating under different reliability constraints.}}
         \label{fig:energy_imprecision_trade_off_users_different_constraints}
     \end{subfigure}
     \hfill
        \caption{\textcolor{black}{FNR evolution and average energy vs. precision trade-off for users operating under different long-term reliability constraints.}}
        \label{fig:heterogeneous_inference_environment}
\end{figure*}

\textcolor{black}{
\section{Additional Results}\label{sec:additional_results}
\subsection{Heterogeneous Edge-Inference Scenarios}
To demonstrate the applicability of the proposed approach to general scenarios, we evaluate the performance of CLO in remote inference scenarios where users operate under diverse reliability constraints. To this end, we consider $K=3$ EDs connected through the multi-hop network architecture depicted in Figure~\ref{fig:multi-hop-network}. The simulation parameters are consistent with those described in Section~\ref{sec:prec_rel_to}, and the frame size is set to $S = 10$ slots. We focus on a binary image segmentation task in which users operate under distinct long-term FNR constraints, namely $r^k = \{0.14, 0.15, 0.20\}$ for ED~1, ED~2, and ED~3, respectively.}

\textcolor{black}{Figure~\ref{fig:time_to_reliability_users_different_constraints} depicts the evolution of the FNR over time for each user connected to the network. It can be observed that the long-term reliability of each user converges to the prescribed target value, thereby demonstrating the capacity of CLO to accommodate users operating under heterogeneous reliability requirements. Conversely, Figure~\ref{fig:energy_imprecision_trade_off_users_different_constraints} illustrates the trade-off between the average network energy consumption and the average precision, evaluated as $1 - F_s(x,\theta)$, experienced by each device. Results have been averaged after $5,{000}$ time slots, at the convergence of the long-term reliability constraint. Due to the intrinsic trade-off between precision and reliability, the average precision degrades as the stringency of the long-term reliability constraint increases.
}

\textcolor{black}{\subsection{Long-term Reliability under Finite Propagation and Estimation Delay}}
\textcolor{black}{We test CLO over the multi-hop network architecture illustrated in Figure~\ref{fig:multi-hop-network}, using the same simulation parameters specified in Section~\ref{sec:prec_rel_to}. We consider a frame  size size $S = 10$ slots, and uniform delay values  $d^k = \{0, 5, 10 \}$ frames are adopted for all users. Figure~\ref{fig:delay_analysis} analyzes the long-term reliability under varying delay conditions. As the delay $d^k$ increases, the violation of the long-term reliability constraint also grows. However, it consistently remains within the theoretical upper bounds, indicated by the dashed lines, which—according to Proposition~\ref{prop:rrc_lyap}—are within an additive error $\mathcal{O}(1/d^k)$ from the ideal case (i.e., when $d^k = 0$). This demonstrates that the signaling overhead due to the computation of the frame loss and the propagation of the updated reliability hyperparameters, do not prevent the algorithm from achieving the target reliability, but only affect the speed of convergence.}

\textcolor{black}{
\begin{figure}
    \centering
    \includegraphics[width=0.85\linewidth]{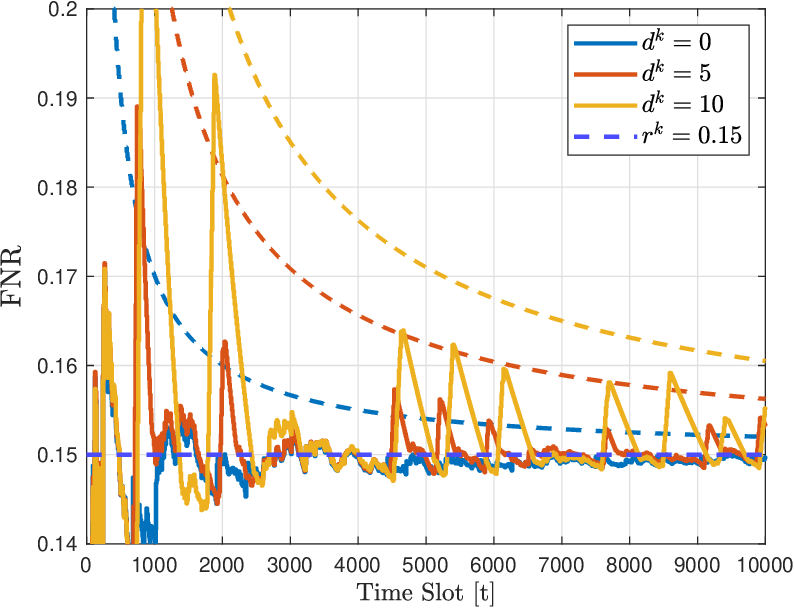}
    \caption{\textcolor{black}{Long-term reliability under different estimation and propagation delay of the CLO hyperparameters update.}}
    \label{fig:delay_analysis}
\end{figure}
}

\textcolor{black}{
\subsection{Impact of the Precision Loss Approximation}\label{sec:precision_loss_approximation_results}
}
\textcolor{black}{To evaluate the performance degradation caused by imprecise precision loss estimation using low-complexity neural networks, we compare the proposed CLO strategy with a \textit{genie-aided resource allocation approach}. In this benchmark, rather than using precision loss approximators, we directly utilize the segmentation networks deployed in the system to guide the resource allocation. While this approach lacks practical applicability, it serves as a meaningful upper bound, allowing us to quantify the degradation introduced by the use of low-complexity approximators in the resource allocation process.}

\textcolor{black}{To this aim, we assess the genie-aided resource allocation policy in the multi-hop network architecture reported in Figure \ref{fig:multi-hop-network}. Figure \ref{fig:energy_precision_trade_off_genie_aided} shows the average energy/precision trade-off reached by the two optimization strategies. The trade-off curves have been obtained simulating CLO for increasing values of the trade-off parameter $\eta$ in \eqref{eq:cost_function} over 10 independent realizations of the task sequence, and averaging the results over the last 1,000 slots, at the convergence of the reliability constraints. Figure \ref{fig:energy_precision_trade_off_genie_aided} testifies that, on average, guiding the resource allocation policy with low-complexity neural network approximators leads to an overall performance loss around the range $1\%-2\%$.
\begin{figure}[h]
    \centering
    \includegraphics[width=0.85\linewidth]{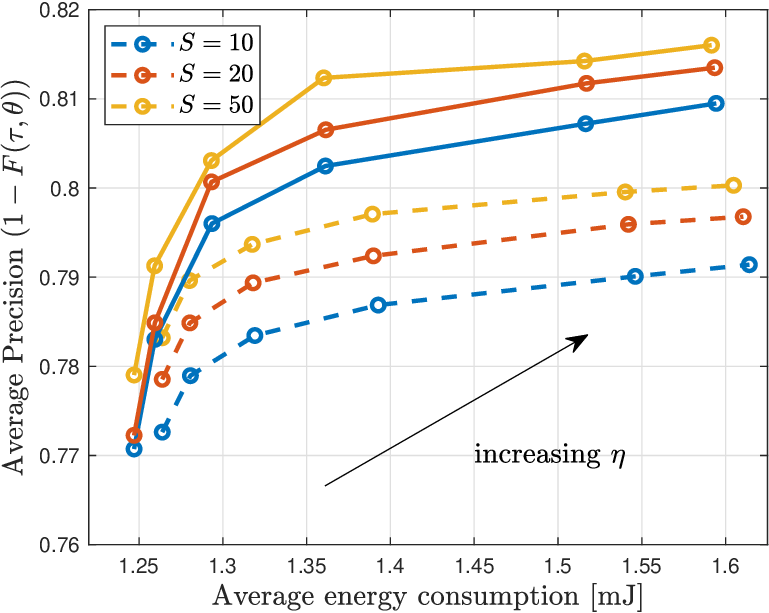}
    \caption{\textcolor{black}{Energy vs. precision trade-off for genie-aided CLO (solid lines) and for CLO driven by precision-loss approximators (dashed lines).}}
    \label{fig:energy_precision_trade_off_genie_aided}
\end{figure}
}

\end{appendices}
\bibliographystyle{ieeetr}
\bibliography{bibliography}

\end{document}